\documentclass[preprint,aps]{revtex4}

\usepackage{epsfig}

\usepackage{float}

\begin{document}

\title{Fitting dwarf galaxy rotation curves with conformal gravity}

\author{James~G.~O'Brien${}^1$ and Philip~D.~Mannheim${}^2$}
\affiliation{${}^1$Department of Applied Mathematics and Sciences\\ Wentworth Institute of Technology\\ Boston, MA 02115, USA
\\{\tt electronic address: obrienj10@wit.edu}\\
${}^2$Department of Physics\\ University of Connecticut\\ Storrs, CT 06269, USA
\\{\tt electronic address: philip.mannheim@uconn.edu}\\ }
\date{October 21, 2011}
\begin{abstract}
We continue our study of the application of the conformal gravity theory to galactic rotation curves. Previously we had studied a varied 111 spiral galaxy sample  consisting of high surface brightness galaxies, low surface brightness galaxies and dwarf galaxies. With no free parameters other than galactic mass to light ratios, we had found that the theory is able to account for the systematics that is observed in the entire set of galactic rotation curves without the need for any dark matter whatsoever. In the present paper we extend our study to incorporate a further 27 galaxies of which 25 are dwarf galaxies and provide updated studies of 3 additional galaxies that had been in the original sample, and again without dark matter find fully acceptable fits, save only for just a few galaxies that we find to be somewhat troublesome. Our current study brings to 138 the number of rotation curves of galaxies that have been accounted for by the conformal gravity theory. Since one of the primary ingredients in the theory is a universal contribution to galactic motions coming from matter exterior to the galaxies, and thus independent of them, our study reinforces one of the central concepts of the conformal gravity studies, namely  that invoking dark matter should be viewed as being nothing more than an attempt to describe global physics contributions in purely local galactic terms.
\end{abstract}

\maketitle

\section{Introduction}
\label{s1}

As a possible alternative to standard Einstein gravity, Weyl introduced conformal gravity in the very early days of general relativity. It is an attractive theory in that it is a pure metric theory of gravity that possesses all of the general coordinate invariance and equivalence principle structure of standard gravity while augmenting it with an additional symmetry, local conformal invariance, in which  the action is left invariant under local conformal transformations on the metric of the form $g_{\mu\nu}(x)\rightarrow e^{2\alpha(x)}g_{\mu\nu}(x)$ with arbitrary local phase $\alpha(x)$. Under such a symmetry the gravitational action is uniquely prescribed to be of the form (see e.g. \cite{Mannheim2006}) 
\begin{equation}
I_{\rm W}=-\alpha_g\int d^4x\, (-g)^{1/2}C_{\lambda\mu\nu\kappa}
C^{\lambda\mu\nu\kappa}
\equiv -2\alpha_g\int d^4x\, (-g)^{1/2}\left[R_{\mu\kappa}R^{\mu\kappa}-(1/3) (R^{\alpha}_{\phantom{\alpha}\alpha})^2\right],
\label{182}
\end{equation}
where 
\begin{equation}
C_{\lambda\mu\nu\kappa}= R_{\lambda\mu\nu\kappa}
-\frac{1}{2}\left(g_{\lambda\nu}R_{\mu\kappa}-
g_{\lambda\kappa}R_{\mu\nu}-
g_{\mu\nu}R_{\lambda\kappa}+
g_{\mu\kappa}R_{\lambda\nu}\right)
+\frac{1}{6}R^{\alpha}_{\phantom{\alpha}\alpha}\left(
g_{\lambda\nu}g_{\mu\kappa}-
g_{\lambda\kappa}g_{\mu\nu}\right)
\label{180}
\end{equation}
is the conformal Weyl tensor and the gravitational coupling constant $\alpha_g$ is dimensionless. 
With the conformal symmetry forbidding the presence of any $\int d^4x\, (-g)^{1/2}\Lambda$ term in the action, the conformal theory has a control over the cosmological constant that the standard Einstein theory does not, and through this control one is able to both address and resolve the cosmological constant problem \cite{Mannheim2009}. Similarly, with the coupling constant $\alpha_g$ being dimensionless, unlike standard gravity conformal gravity is renormalizable; and with it having been shown \cite{Bender2008,Mannheim2009} to be unitary at the quantum level, the theory is offered \cite{Mannheim2011} as a consistent theory of quantum gravity in four spacetime dimensions.

With the conformal theory being a consistent, renormalizable quantum theory at the microscopic level, then just as with electrodynamics,  one is assured that its macroscopic classical predictions are reliable and will not be ruined by quantum corrections. Consequently, application of the theory to astrophysical phenomena allows one to test the theory. Early work in this direction was provided in \cite{Mannheim1997} where the theory was used to fit the rotation curves of a set of 11 spiral galaxies, with the mass to light ratios ($M/L$) of the luminous optical disk of each galaxy being the only free parameters, and with no dark matter being required. More recently \cite{Mannheim2010a,Mannheim2010b} a systematic, broad-based study of the rotation curves of a varied set of 111 galaxies (consisting of high surface brightness galaxies, low surface brightness galaxies and dwarf galaxies) was made (the set included the original 11 galaxies and updates where available), and again acceptable fitting was obtained with the optical disk mass to light ratios of each galaxy being the only free parameters, and again with no dark matter being required. In this paper we extend the study of \cite{Mannheim2010a,Mannheim2010b} to a set of 30 galaxies that contains 20 of the dwarf galaxies described in \cite{Swaters2009}, 2 low surface brightness galaxies (LSB), 5 of the dwarf galaxies described in \cite{Oh2011}, and updates of 3 galaxies (2 dwarf galaxies and an LSB) that were in the 111 galaxy sample, and report findings of the same quality as before. With this latest study, the sample of galactic rotation curves  that can be accounted for by the conformal theory now runs to 138. 

\section{General Formalism}
\label{s2}

The general formalism for applying the conformal theory to galactic rotation curves has been described in detail in \cite{Mannheim2010a,Mannheim2010b}, and we recall the main results. For the Weyl action $I_{\rm W}$ given in  (\ref{182}), functional variation with respect to the metric $g_{\mu\nu}(x)$ generates a gravitational equation of motion of the form \cite{Mannheim2006} 
\begin{equation}
4\alpha_g W^{\mu\nu}=4\alpha_g\left[
2C^{\mu\lambda\nu\kappa}_
{\phantom{\mu\lambda\nu\kappa};\lambda;\kappa}-
C^{\mu\lambda\nu\kappa}R_{\lambda\kappa}\right]=4\alpha_g\left[W^{\mu
\nu}_{(2)}-\frac{1}{3}W^{\mu\nu}_{(1)}\right]=T^{\mu\nu},
\label{188}
\end{equation}
where
\begin{eqnarray}
W^{\mu \nu}_{(1)}&=&
2g^{\mu\nu}(R^{\alpha}_{\phantom{\alpha}\alpha})          
^{;\beta}_{\phantom{;\beta};\beta}                                              
-2(R^{\alpha}_{\phantom{\alpha}\alpha})^{;\mu;\nu}                           
-2 R^{\alpha}_{\phantom{\alpha}\alpha}
R^{\mu\nu}                              
+\frac{1}{2}g^{\mu\nu}(R^{\alpha}_{\phantom{\alpha}\alpha})^2,
\nonumber\\
W^{\mu \nu}_{(2)}&=&
\frac{1}{2}g^{\mu\nu}(R^{\alpha}_{\phantom{\alpha}\alpha})   
^{;\beta}_{\phantom{;\beta};\beta}+
R^{\mu\nu;\beta}_{\phantom{\mu\nu;\beta};\beta}                     
 -R^{\mu\beta;\nu}_{\phantom{\mu\beta;\nu};\beta}                        
-R^{\nu \beta;\mu}_{\phantom{\nu \beta;\mu};\beta}                          
 - 2R^{\mu\beta}R^{\nu}_{\phantom{\nu}\beta}                                    
+\frac{1}{2}g^{\mu\nu}R_{\alpha\beta}R^{\alpha\beta}.
\label{108}
\end{eqnarray}                                 
For the case of a static, spherically symmetric geometry,  it was shown in  
\cite{Mannheim1989,Mannheim1994} that without loss of generality, the exact, all-order classical line element could be brought to the form 
\begin{equation}
ds^2=-B(r)c^2dt^2+\frac{dr^2}{B(r)}+r^2d\theta^2+r^2\sin^2\theta~d\phi^2,
\label{108a}
\end{equation}
and that in terms of this line element the exact fourth-order equation of motion given in (\ref{188}) could be reduced to the remarkably simple  fourth-order Poisson equation form
\begin{equation}
\frac{3}{B(r)}(W^{0}_{\phantom{0}0}-W^{r}_{\phantom{r}r})=\nabla^4B=\frac{d^4B}{dr^4}+\frac{4}{r}\frac{d^3B}{dr^3}=\frac{3}{4\alpha_gB(r)}(T^{0}_{\phantom{0}0}-T^{r}_{\phantom{r}r})=f(r)
\label{E2}
\end{equation}
without any approximation whatsoever. 

Solutions to (\ref{E2}) are of the form
\begin{eqnarray}
B(r)&=& -\frac{r}{2}\int_0^r
dr^{\prime}\,r^{\prime 2}f(r^{\prime})
-\frac{1}{6r}\int_0^r
dr^{\prime}\,r^{\prime 4}f(r^{\prime})
\nonumber\\
&&-\frac{1}{2}\int_r^{\infty}
dr^{\prime}\,r^{\prime 3}f(r^{\prime})
-\frac{r^2}{6}\int_r^{\infty}
dr^{\prime}\,r^{\prime }f(r^{\prime}) +\hat{B}(r),
\label{P129}
\end{eqnarray}                                 
where $\hat{B}(r)$ is the general solution to $\nabla^4\hat{B}(r)=0$. With $B^{\prime}(r)$ evaluating to 
\begin{eqnarray}
B^{\prime}(r)&=& -\frac{1}{2}\int_0^r
dr^{\prime}\,r^{\prime 2}f(r^{\prime})
+\frac{1}{6r^2}\int_0^r
dr^{\prime}\,r^{\prime 4}f(r^{\prime})
-\frac{r}{3}\int_r^{\infty}
dr^{\prime}\,r^{\prime }f(r^{\prime}) +\hat{B}^{\prime}(r),
\label{P129a}
\end{eqnarray}                                 
and with the non-relativistic gravitational force being given by $B^{\prime}(r)c^2/2$, we recognize two classes of contribution to the force, a local contribution due to the first two terms  in (\ref{P129a}) and a global one due to the last two terms. 

For a localized system such as a star whose source function $f(r)$  is restricted to its interior $0\leq r \leq r_0$ region, only the first two integrals in (\ref{P129}) contribute to $B(r>r_0)$ and yield
\begin{equation}
B(r>r_0)=\frac{2\beta}{r}+\gamma r,
\label{P130}
\end{equation}
where
\begin{equation}
2\beta=\frac{1}{6}\int_0^{r_0}\,dr^{\prime}\,r^{\prime 4}f(r^{\prime}),\qquad \gamma= -\frac{1}{2}\int_0^{r_0}dr^{\prime}\,r^{\prime 2}f(r^{\prime}).
\label{P131}
\end{equation}
With  the luminous material in a spiral galaxy typically being distributed with a surface brightness of the form $\Sigma(R)=e^{-R/R_0}L/2\pi R_0^2$ where $L$ is the total luminosity, and with the potential produced by a single star being of the form $V^{*}(r)=-\beta^{*}c^2/r+\gamma^{*} c^2 r/2$ as per (\ref{P130}), the local galactic potential $V_{\rm LOC}(r)$ is readily computed \cite{Mannheim2006}, with the contribution of the material in a galaxy to rotation velocities then being given by the very compact formula
\begin{eqnarray}
v_{\rm LOC}^2(R)&=&
\frac{N^*\beta^*c^2 R^2}{2R_0^3}\left[I_0\left(\frac{R}{2R_0}
\right)K_0\left(\frac{R}{2R_0}\right)-
I_1\left(\frac{R}{2R_0}\right)
K_1\left(\frac{R}{2R_0}\right)\right]
\nonumber \\
&+&\frac{N^*\gamma^* c^2R^2}{2R_0}I_1\left(\frac{R}{2R_0}\right)
K_1\left(\frac{R}{2R_0}\right),
\label{P136}
\end{eqnarray} 
where $N^*M_{\odot}=M=(M/L)L$ is the total mass of the galaxy.

Unlike the standard second-order gravitational Poisson equation where only local material within a galaxy contributes to the gravitational force, in the fourth-order conformal case there are contributions coming from material outside the galaxy as well. These global contributions  are due to the homogeneous cosmological background and the inhomogeneities in it. With the $\beta^*$-dependent term in (\ref{P136}) being the standard galactic luminous Newtonian term, it is these global contributions together with the $\gamma^*$-dependent term in (\ref{P136}) that are to replace dark matter, with dark matter potentially being nothing more than an attempt to describe global physics contributions in purely local galactic terms.

Since $W_{\mu\nu}$ given in (\ref{188}) is zero in solutions that obey $\nabla^4\hat{B}(r)=0$, and since $W_{\mu\nu}$ vanishes in the conformal to flat  Robertson-Walker (RW) geometry, the $\hat{B}(r)$ contribution is due to the background cosmology itself. For the specific form of the contribution,  we recall \cite{Mannheim1989,Mannheim1997} that a coordinate transformation of the form
\begin{equation}
\rho=\frac{4r}{2(1+\gamma_0r)^{1/2}+2 +\gamma_0 r},\qquad r= \frac{\rho}{(1-\gamma_0\rho/4)^2},\qquad \tau=\int dt\,a(t)
\label{P133}
\end{equation}                                 
effects the metric transformation
\begin{eqnarray}
&&-(1+\gamma_0r)c^2dt^2+\frac{dr^2}{(1+\gamma_0r)}+r^{ 2}d\theta^2+r^{2} \sin^2\theta d\phi^2
\nonumber \\
&&=\frac{1}{a^2(\tau)}\left(\frac{1+\gamma_0\rho/4}
{1-\gamma_0\rho/4}\right)^2
\left[-c^2d\tau^2+\frac{a^2(\tau)}{[1-\gamma_0^2\rho^2/16]^2}
\left(d\rho^2+\rho^2d\theta^2+\rho^2 \sin^2\theta d\phi^2\right)\right].
\label{P134}
\end{eqnarray} 
With the bracketed term on the right-hand side of  (\ref{P134}) being recognized as an RW geometry with an expressly negative definite $K=-\gamma_0^2/4$, we see that in the rest frame of a galaxy, a $K<0$ cosmological background acts as a universal linear potential $\gamma_0$ whose strength is independent of the galaxy of interest. Since as noted in \cite{Mannheim2006} conformal gravity precisely produces such a $K<0$ RW background, its effect on galactic rotation curves is to produce an effective universal linear potential term.

As regards the contribution due to inhomogeneities, it was noted in \cite{Mannheim2010a,Mannheim2010b} that if the exterior $f(r)$ distribution starts at some typical cluster radius $r_{\rm clus}$, then in the $r_0\leq r \leq r_{\rm clus}$ region the contribution of the fourth integral in (\ref{P129}) acts like a quadratic de Sitter like potential with a strength 
\begin{equation}
\kappa =\frac{1}{6}\int_{r_{\rm clus}}^{\infty} dr^{\prime}\,r^{\prime }f(r^{\prime}).
\label{P132}
\end{equation}
Since the integral in (\ref{P132}) is independent of the galaxy of interest, its contribution is universal for all galaxies, and since it is due to all the inhomogeneous material in the Universe, it would equally act on galaxies that are within clusters as well.

Finally, for weak gravity, and on scales $r < r_{\rm clus}$,  we can augment (\ref{P136}) with the linear and quadratic potential terms, and obtain a total contribution $v_{{\rm TOT}}^2(R)$ to the centripetal velocity of the form 
\begin{equation}
v^2_{\rm TOT}(R)=v^2_{\rm LOC}(R)+\frac{\gamma_0 c^2R}{2}-\kappa c^2R^2,
\label{E20}
\end{equation}                                 
with associated asymptotic limit 
\begin{equation}
v_{{\rm TOT}}^2(R) \rightarrow \frac{N^*\beta^*c^2}{R}+
\frac{N^*\gamma^*c^2R}{2}+\frac{\gamma_0c^2R}{2}-\kappa c^2R^2.
\label{E21}
\end{equation} 
Equation (\ref{E20}) is our key result, with the mass to light ratio  of the luminous disk in $v^2_{\rm LOC}(R)$ being the only free parameter in any given galaxy, and with everything else being universal.

In \cite{Mannheim1997,Mannheim2010a,Mannheim2010b}   (\ref{E20}) was used to fit the galactic rotation curve data of a sample of 111 galaxies, and good fits were found, with the three universal parameters being given by
\begin{equation}
\gamma^*=5.42\times 10^{-41}~{\rm cm}^{-1},\qquad \gamma_0=3.06\times
10^{-30}~{\rm cm}^{-1},\qquad \kappa =9.54\times 10^{-54}~{\rm cm}^{-2}.
\label{E18}
\end{equation} 
The value obtained for $\gamma^*$ entails that the linear potential of the Sun is so small that there are no modifications to standard solar system phenomenology, with the values obtained for $N^*\gamma^*$, $\gamma_0$ and $\kappa$ being such that one has to go all the way to galactic scales before their effects can become as big as the Newtonian contribution. The value obtained for $\gamma_0$ shows it to indeed be of cosmological magnitude, with the value of $\kappa$ being a typical 100 Mpc inhomogeneity scale, just as desired. Armed with the above analysis and the values of the 3 universal parameters as given in (\ref{E18}), we now proceed to apply (\ref{E20}) to the 30 galaxy sample.

\section{Conformal Gravity Fitting to the 30 Galaxy Sample}
\label{s4}

The sample of galaxies we study is composed of a 24 spiral galaxy sample and a 6 galaxy sample.
The 24 galaxy sample consists of 23 of the 27 galaxies that were studied in \cite{Swaters2010} as augmented by UGC 12732, with 6 dwarf galaxies being taken from the THINGS collaboration as reported in \cite{Oh2011}. Of the 24 galaxies in the \cite{Swaters2010} sample 21 of them (UGC 731, UGC 3371, UGC 4173, UGC 4325, UGC 4499, UGC 5414, UGC 5721, UGC 7232, UGC 7323, UGC 7399, UGC 7524, UGC 7559, UGC 7577, UGC 7603, UGC 8490, UGC 9211, UGC 11707, UGC 11861, UGC 12060, UGC 12632 and UGC 12732) belong to the late-type dwarf galaxy sample described in \cite{Swaters2009}, 2 are LSB galaxies to which we had not previously applied conformal gravity fitting (F568-V1, F574-1), with the remaining galaxy being the LSB galaxy UGC 5750. Even though UGC 4325 and UGC 5750 had both been contained in our previously studied 111 galaxy sample, the data we use for them now are sufficiently different to warrant their inclusion here. (The remaining 4 galaxies studied in \cite{Swaters2010} (F583-1, F583-4, UGC 5005, UGC 6446) were all contained in our 111 galaxy sample and are not included here as there is no significant change in either the data or in our fits to them.) For the 21 dwarf galaxies we use the rotation curve data reported in \cite{Swaters1999} and \cite{Swaters2009}. For  F568-V1 and  F574-1 we use the rotation curve data reported in \cite{Swaters2003}. For UGC 5750  we use the rotation curve data reported in \cite{deBlok2001}, data that extend to the much larger distances of significance to conformal gravity than had been considered in \cite{Mannheim2010b}. For the fitting we take B-band absolute luminosities and R-band optical disk scale lengths from \cite{Swaters2002a}, \cite{Swaters2002b} and \cite{Swaters2010} and references therein (using B-R=0.8 to extract the B-band luminosity of UGC 11861 from the R-band luminosity value listed in \cite{Swaters2002b}), except that for UGC 5750 we take the B-band luminosity from \cite{deBlok1997}. For the 24 galaxies we take HI gas masses from \cite{Swaters2002b} and \cite{Swaters2010}. 

The 6 galaxy sample presented in  \cite{Oh2011} consists of UGC 3851 (NGC 2366, DDO 42), UGC 4305 (Holmberg II, DDO 50), UGC 4459 (DDO 53), UGC 5139 (Holmberg I, DDO 63), UGC 5423 (M81dwB ), and UGC 5666 (IC2754, DDO 81). The galaxy UGC 5666 had been included in our earlier 111 galaxy study, but the data have changed sufficiently to warrant its being included here. (The remaining galaxy studied in \cite{Oh2011} (DDO 154) was also contained in our 111 galaxy sample and is not included here as there has been no significant change in the data since then.) For the 6 dwarf galaxies we use the rotation curve data reported in \cite{Oh2011}, and take B-band absolute luminosities and HI gas masses from \cite{Walter2008}.  For UGC 3851, UGC 4305  and UGC 5139 we take R-band disk scale lengths from \cite {Swaters2002b}, for  UGC 4459 and UGC 5423 we follow \cite{Oh2011} and take disk scale lengths to be 5 times the $z_0$ scale heights they report, and for UGC 5666 we take the disk scale length from \cite{Pasquali2008}.

In our theory the place where there is the most sensitivity to parameters is in the adopted distances to the individual galaxies, since the parameters $\gamma^*$, $\gamma_0$ and $\kappa$ that appear in (\ref{E20}) are given as absolute quantities. Moreover, for dwarf galaxies the parameter $N^*$ is usually so small that the $N^*\gamma^*$-dependent terms in (\ref{E20}) and (\ref{E21}) cannot compete with the $\gamma_0$ term, while the $N^*\beta^*$-dependent terms in (\ref{E20}) and(\ref{E21}) are falling off at the largest distances in galaxies where the mass discrepancy problem is the most severe, so they do not compete with the $\gamma_0$ term either. In addition, except for only a few cases, the dwarf galaxy data do not go out far enough from galactic centers for the $\kappa$ term contribution to be that significant. Hence the $\gamma_0c^2R/2$ term in (\ref{E20}) is the most relevant for our study, and it requires a good determination of the adopted distance to each galaxy.

To establish a common baseline for determining the needed adopted distances, for all the galaxies in our sample we follow our earlier studies \cite{Mannheim2010a,Mannheim2010b} and use the distances listed in the NASA/IPAC Extragalactic Database (NED). In this database distances are obtained either via direct visual measurements (typically Cepheids or the Tully-Fisher relation) or indirectly via redshift measurements. For the directly determined distances a world average mean value and its one standard deviation uncertainty are listed. The redshift-based determinations depend on how one models both the peculiar velocity with respect to the Hubble flow of the Milky Way Galaxy and the peculiar velocity of the galaxy of interest. For definitiveness, for redshift-based distance determinations (as needed for F568-V1, F574-1, and UGC 5750) we have opted to use the mean values associated with galactocentric  distance determinations. 

In Tables 1 and 2 we list the mean NED values for adopted distances to the 30 galaxies in our sample, and assemble all the optical and HI gas input data as determined at the NED mean distances.  In Tables 1 and 2 we also list the values $R_{\rm last}$ of the locations of the last data points in each of the 30 galaxies as determined at the NED mean distances. In the fitting we multiply the HI gas masses given in Tables 1 and 2 by 1.4 to account for primordial Helium.

A second place in our theory where there is sensitivity to parameters is in the values of galactic inclinations that are extracted from the rotation curve data, since changes in inclination entail changes in the extracted values of rotational velocities and their errors, with the $\gamma_0c^2R/2$ term needing to account for whatever the extracted values are. Specifically, with the inclination angle being the angle between the plane of the galactic disk and our line of sight, a measured Doppler shift $v({\rm meas})$ is actually a measurement of the projection $v(i)\sin i$ at inclination angle $i$ along our line of sight (so that for edge-on galaxies $v(90^\circ)=v({\rm meas})$). At two different inclinations the associated velocities would thus be related as $v({\rm meas})=v(i_1)\sin i_1=v(i_2)\sin i_2$, with a decrease in inclination leading to an increase in inferred rotation velocity. By the same token, since the gas mass is determined by the number $N({\rm  meas})$ of 21 cm photons that reach us, the number of hydrogen atom emitters $N(i)$ at inclination $i$ is determined by HI column depth along our line of sight according to $N({\rm meas})=N(i)/\cos i$, being minimal for face-on galaxies where $N(0^{\circ})=N({\rm meas})$. At two different inclinations the inferred HI gas masses would  be related as $M_{\rm HI}(i_1)/\cos i_1=M_{\rm HI}(i_2)/\cos i_2$, with a decrease in inclination leading to an increase in inferred HI gas mass. The rotation curve data that are displayed in Figs. 1 and 2 and the HI gas masses that are given in Tables 1 and 2 were obtained using the inclinations given in \cite{Swaters2010} and \cite{Oh2011} as listed in Tables 1 and 2.

Since the HI gas distributions of the galaxies in our sample are found to extend well beyond the associated optical disk regions \cite{Swaters2002b}, just as in \cite{Mannheim2010a,Mannheim2010b} we have approximated the gas profiles as single exponential disks with scale lengths equal to 4 times those of the corresponding optical disks.  For our purposes here this is very convenient as it allows us to use (\ref{P136}) as is for the gas contribution to $v_{\rm LOC}^2(R)$. Moreover, in the region in a galaxy where $R$ is much greater than the gas scale length, the asymptotic (\ref{E21}) will then hold with $N^*$ now being understood to connote the stars and gas masses combined. Since for dwarf galaxies such a combined $N^*\gamma^*c^2R/2$ term does not compete in (\ref{E21}) with the $\gamma_0c^2R/2$ term while the $N^*\beta^*c^2/R$ term is falling off, the $\gamma_0c^2R/2$ term will dominate in the outer region regardless of what specific values for the gas scale lengths we might actually use. With the use of single exponential disks for the gas distributions, we  can then directly monitor the outer regions of rotation curves, the regions where the galactic mass discrepancy problem is at its most pronounced.  As regards the inner rotation curve regions, even though there could be some sensitivity to the forms we might use for the gas distributions, in practice we found fits of essentially identical quality using a variety of smaller gas scale length to optical scale length ratios, and in this paper only report the fits we obtain in which the gas to optical scale length ratios are taken to be equal to 4 in each galaxy.

To get an immediate sense of the expectations of the conformal theory, we fit the rotation curves of the 30 galaxies in our sample using the input data listed in Tables 1 and 2 as is, without regard to  any uncertainties in adopted distances or galactic inclinations.  In Figs. 1  and 2 we present  the resulting rotational velocities with their quoted errors (in ${\rm km}~{\rm sec}^{-1}$) for all of the 30 galaxies as plotted as functions of radial distances from galactic centers (in ${\rm kpc}$). For each galaxy we have exhibited the contribution due to the luminous (stars plus gas) Newtonian term alone (dashed curve), the contribution from the two linear terms alone (dot dashed curve), the contribution from the two linear terms and the quadratic term combined as per (\ref{E20}) (dotted curve), with the full curve showing the total contribution.  

In Tables 1 and 2 we list the fitted optical disk masses and mass to light ratios that we obtained in the conformal gravity fitting. As described in \cite{Mannheim2010b}, we constrained the fits so that the mass to light ratios would not be less than 0.2 nor larger than 10.0. The mass to light ratios that we obtain are typical of the values that one ordinarily obtains in galactic rotation curve fitting, and are reasonably close to the $M/L$ ratio in the local solar neighborhood. 

As we see, use of the tightly constrained (\ref{E20}) and the tightly constrained input values given in Tables 1 and 2 captures the general trend of  the data to a remarkably good degree, and does so without needing any dark matter whatsoever. The fact that our fits work so well even though the velocities are dominated by a single galaxy-independent universal $\gamma_0c^2R/2$ term in (\ref{E20}) is thus a quite noteworthy achievement for our theory. 

In the fitting the only real concern we have is for UGC 5721, UGC 7577 and UGC 4305, as their overall normalizations are not well accounted for. Improved fitting for these 3 galaxies can be obtained by using NED distances as modified by one standard deviation distance uncertainties, and by using different inclination angles. For UGC 5721 we found it advantageous to increase the adopted  distance by one standard deviation and increase the inclination by 10 degrees. In addition, for UGC 5721 we took advantage of the uncertainty in the disk scale length reported in \cite{Spano2008}, and also incorporated an optical disk thickness correction of the form $f(z)={\rm sech}^2(z/z_0)/2z_0$ with $z_0=R_0/5$ as per the formalism given in  \cite{Mannheim2006}. For UGC 7577 we found it advantageous to decrease the adopted  distance by one standard deviation, and on adding (in quadrature) an overall asymmetric drift error of 3 ${\rm km}~{\rm sec}^{-1}$ that was reported in \cite{Swaters2009} but not allowed for  in the velocity errors shown in Fig. 1, we obtained good fitting if we reduced the  inclination by 15 degrees. [Without this asymmetric drift correction we could obtain a comparable fit (not shown) with an inclination reduction of 20 degrees.] For UGC 4305 we found it advantageous to decrease the adopted  distance by one standard deviation and reduce the inclination by 5 degrees, to then be midway between the inclinations of  50 degrees and 40 degrees that were respectively reported in  \cite{Oh2011} and \cite{Swaters2009}.

For  4 other of the galaxies we also have concerns, though they are mild, and making analogous adjustments proved beneficial. For UGC 7399 we found it advantageous to increase the adopted  distance by one standard deviation. For UGC 4173 we found it advantageous to reduce the inclination angle by 5 degrees, for UGC 8490 we increased the inclination by 10 degrees, and for UGC 3851 we increased the adopted distance by one standard deviation. With all the above changes we obtained the fits given in Fig. 3, with the associated parameters being listed in Table 3. The fitting to the galaxy UGC 5666 is of interest in that while we had found it beneficial to increase its adopted distance by one standard deviation in our earlier study of it \cite{Mannheim2010b}, with the new data we can now use the NED mean distance as is.

With the adjustments we have made, the only galaxy for which we still have some difficulty is UGC 5721, since even though the conformal theory could readily reproduce the rotation curve in either the outer region or the inner region alone (not shown), the presented eyeball fit  represents the result of trying to accommodate both regions simultaneously. However, it was noted in \cite{Swaters2009} that for this galaxy the inner rotation curve points are uncertain because of insufficient angular resolution, while in \cite{Swaters2002a} it was noted that that there is evidence for twisting isophotes in the central region, to thus suggest additional structure in the inner region of this galaxy. We have not taken any such inner region structure into consideration in our fitting. Apart from this one inner region concern and apart from the use of a somewhat large  inclination shift for UGC 7577, the conformal theory is otherwise found to account for the rotation curve data of the entire 30 galaxy sample to a quite remarkable degree.

\section{General Comments}
\label{s5}

The cumulative set of 138 galaxy fits presented here and in \cite{Mannheim2010a,Mannheim2010b} is noteworthy in that the universal $\gamma_0$ and $\kappa$ terms have no dependence on individual galactic properties whatsoever and yet have to work in every single case. Our fits are also noteworthy in that we have captured the essence of the rotation curve data even though we have imposed some rather strong constraints on the input parameters. For adopted distances in most cases we have used NED mean values. We have not used actual surface brightness distributions or actual gas profiles but have treated these distributions simply as single exponentials. On the theoretical side our fits are noteworthy in that (\ref{E20}) is not simply a phenomenological or empirical formula that was extracted solely from consideration of the systematics of galactic rotation curves. Rather, (\ref{E20}) was explicitly derived from first principles in a fundamental, uniquely prescribed metric-based theory of gravity, namely conformal gravity. Moreover, conformal gravity  itself was not even advanced for the purposes of addressing the dark matter problem. Rather, before it was known what its static, spherically symmetric solutions might even look like, it was advanced by one of us \cite{Mannheim1990} simply because it had a symmetry that could control the cosmological constant. Our fitting is thus quite non-trivial.

Since the conformal gravity fits do capture the essence of the data, it is important to ask how it is that a theory with so few adjustable parameters is actually able to do so. As was already noted in \cite{Mannheim2010a,Mannheim2010b} for the 111 galaxy sample, those data are such that the value of the quantity $(v^2/c^2R)_{\rm last}$ as measured at the last data point for each of those galaxies is close in value to  $\gamma_0$ across the entire 111 galaxy sample. As we see from the last column in Table 1, this very same last data point near universality is also possessed by the additional galaxies that we study here, with all of the $(v^2/c^2R)_{\rm last}$ values clustering close to the numerically extracted universal value for the parameter $\gamma_0$.

With regard to the near universality of $(v^2/c^2R)_{\rm last}$, we should note that this is an empirical property of the raw data themselves. Moreover, while there may be some uncertainties in the adopted distances to the galaxies, such uncertainties are never more than a factor of two or so. With the velocities being uncertain to no more than 10 to 20 per cent or so, the near universality of $(v^2/c^2R)_{\rm last}$ is thus a genuine property of the data. It should thus be regarded as an important empirical clue for galactic dynamics. 

Apart from conformal gravity, Milgrom's MOND theory \cite{Milgrom1983} and Moffat's  MSTG theory \cite{Moffat2005} equally succeed in explaining galactic rotation curves without invoking dark matter. Each of these particular theories has an assumed universal parameter ($a_0/c^2=1.33\times 10^{-29}~{\rm cm}^{-1}$ for MOND, and $G_0M_0/r_0^2c^2=7.67\times 10^{-29}~{\rm cm}^{-1}$ for MSTG), and in consequence each is able to account for the data. Now these two theories and the conformal theory predict differing behaviors at larger distances, As discussed in \cite{Mannheim2010b}, the MOND theory typically requires asymptotic flatness of rotation curves, the MSTG theories requires Kepler behavior at large distances, while the conformal theory requires a fall in rotation velocities to zero at $R\sim \gamma_0/\kappa$, and a thus finite size to galaxies. Given such differing behaviors, it should eventually be possible to phenomenologically distinguish between these various theories.

It is important to recognize that for conformal gravity fitting (and likewise for MOND and MSTG) the only input one needs is the luminous matter distribution. Then, with only one free parameter per galaxy (viz. the galactic $M/L$ ratio) rotation velocities are completely determined; and as Tables 1 and 2 show, by and large the $M/L$ ratios are all found to be of order the local solar 
neighborhood $M/L$ ratio, just as one would want. Moreover, since the $M/L$ ratios are effectively determined by the inner rotation curve region data alone, the linear and quadratic terms being at their smallest there, the outer region fitting is then essentially parameter free. 

It is important to contrast conformal gravity fitting with dark matter fitting to galactic rotation data. For dark matter fits one first needs to know the velocities so that one can then ascertain the needed amount of dark matter, i.e. in its current formulation dark matter is only a parametrization of the velocity discrepancies that are observed and is not a prediction of them. Even with the freedom to treat galactic mass to light ratios as free parameters, dark matter theory has yet to develop to the point where it is able to determine rotation curve velocities from a knowledge of the luminous matter distribution alone. Nor is it currently able to provide an explanation for the near universality that is found for $(v^2/c^2R)_{\rm last}$. Thus dark matter theories, and in particular those theories that produce dark matter halos in the early Universe (such as the NFW theory \cite{Navarro1996}), are currently unable to make an a priori determination as to which halo is to go with which particular luminous matter distribution, and need to fine-tune halo parameters to luminous parameters galaxy by galaxy. 

The lack to date of any such underlying universal structure or of any explanation for the near universality of $(v^2/c^2R)_{\rm last}$ is more than just a challenge to dark matter theory,  it runs counter to the very motivation that led to the acceptance of Newtonian gravity in the first place. Specifically, the great appeal of Newtonian gravity was not just that it explained planetary orbits via an inverse square force, but that it did so through the use of a solar potential  that possessed just one universal parameter, viz. the solar  Schwarzschild radius $M_{\odot}G/c^2$, that was to control the orbits of all the planets in the solar system. If universal Newtonian gravity is to continue to hold on galactic distance scales, then galactic orbits (and particularly those beyond the galactic optical disk region) should also be explainable in terms of just one parameter, viz. the galactic Schwarzschild radius $N^*M_{\odot}G/c^2$. Thus, if dark matter is to be the explanation, then both the dark to luminous matter mass ratio and the relation between the spatial distributions of dark and luminous matter  should be universal to all galaxies, with rotation curves then being determinable from a knowledge of the luminous mass of the galaxy and its spatial distribution alone. It is the lack of any such universality in dark matter theory that leads to the need to fine-tune its halo parameters galaxy by galaxy. In contrast, no such fine-tuning shortcomings appear in conformal gravity, and if standard gravity is to be the correct  description of gravity, then a universal formula akin to the one given in (\ref{E20}) would need to be derived by dark matter theory. However, since our study establishes that global physics has an influence on local galactic motions, the invoking of dark matter in galaxies could potentially be nothing more than an attempt to describe global physics effects in purely local galactic terms.

The authors wish to thank Dr.~R.~A.~Swaters and Dr. S-H. Oh for helpful communications, and especially for providing their galactic data bases, and wish to thank Dr.~S.~S.~McGaugh for helpful communications. This research has made use of the NASA/IPAC Extragalactic Database (NED) which is operated by the Jet Propulsion Laboratory, California Institute of Technology, under contract with the National Aeronautics and Space Administration. 

\phantom{}

{}

\begin{table}[ht!]
\caption{Properties of the 24  Galaxy Sample}
\centering
\begin{tabular}{l c c c c c c c c c c} 
\hline\hline
\phantom{00}Galaxy\phantom{0}&\phantom{00}Distance  & $L_{\rm  B}$ & ~~$i$~~& $(R_0)_{\rm disk}$  & $R_{\rm last} $ &  $M_{\rm HI} $ & $M_{\rm disk}$ &  $ 
(M/L_{\rm B}) _{\rm disk}$ & $(v^2 / c^2 R)_{\rm last}$  \\  
&   (Mpc)  &  $(10^{9}{\rm L}_{\odot}^{\rm B})$  & $~~^{\circ}~~$ & (kpc) & (kpc) & {$(10^{9} M_\odot)$} & {$(10^{9}
M_\odot)$} & ({$M_{\odot}/L_{\odot}^{\rm B}$}) & {$(10^{-30}\texttt{cm}^{-1})$} \\
\hline

F568-V1\phantom{0} &78.20& \phantom{0}2.15 &40& 3.11 & 17.07 & 2.32 & 16.00 & 7.45 &
2.95 \\

F574-1\phantom{00} &94.10& \phantom{0}3.42 &65& 4.20 & 13.69 & 3.31 & 14.90 & 4.35 &
2.77  \\

UGC \phantom{00}731&11.80& \phantom{0}0.69   &57& 2.43 & 10.30 & 1.61 & \phantom{0}3.21 & 4.63 &
1.91  \\

UGC \phantom{0}3371 &18.75& \phantom{0}1.54  &49& 4.53 & 15.00 & 2.62 & \phantom{0}4.49& 2.91 &
1.78  \\

UGC \phantom{0}4173 &16.70& \phantom{0}0.33   &40& 4.43 & 12.14 & 2.09 & \phantom{0}0.07 & 0.20 &
0.96  \\

UGC \phantom{0}4325 &11.87& \phantom{0}1.71   &41& 1.92 & \phantom{0}6.91& 1.04 & \phantom{0}6.51 & 3.82 &
4.37  \\

UGC \phantom{0}4499 &12.80& \phantom{0}1.01  &50& 1.46 & \phantom{0}8.38 & 1.15 & \phantom{0}1.80& 1.79 &
2.37 \\

UGC \phantom{0}5414 &\phantom{0}9.40& \phantom{0}0.49   &55& 1.40 & \phantom{0}4.10 & 0.57 & \phantom{0}1.13 & 2.29&
3.31\\

UGC \phantom{0}5721 &\phantom{0}6.70& \phantom{0}0.37   &61& 0.46 & \phantom{0}7.42 & 0.66 & \phantom{0}1.03 & 2.76 &
3.00\\

UGC \phantom{0}5750 &56.10& \phantom{0}4.72   &64& 5.60 & 21.77 & 1.00 & \phantom{0}3.68 & 0.78 &
1.03\\

UGC \phantom{0}7232 &\phantom{0}3.14& \phantom{0}0.08  &59& 0.30 & \phantom{0}0.91& 0.06 & \phantom{0}0.14 &1.76 &
7.64\\

UGC \phantom{0}7323 &\phantom{0}7.90& \phantom{0}2.39 &47&  2.13 & \phantom{0}5.75 & 0.70 & \phantom{0}4.19 &1.75 &
4.59\\

UGC \phantom{0}7399 &14.83& \phantom{0}1.67 &55& 1.40 & 19.42 & 2.31 & \phantom{0}5.90 &3.54 &
2.19\\

UGC \phantom{0}7524 &\phantom{0}4.12& \phantom{0}1.37  &46&  3.02 & \phantom{0}9.29 & 1.34 & \phantom{0}5.29 &3.86 &
2.67\\

UGC \phantom{0}7559 &\phantom{0}4.20&\phantom{0}0.04  &61&   0.87 & \phantom{0}2.75 & 0.12 & \phantom{0}0.05 &1.32 &
1.43\\

UGC \phantom{0}7577 &\phantom{0}3.03& \phantom{0}0.10  &63&  0.73 & \phantom{0}1.98 & 0.06 & \phantom{0}0.02 &0.20 &
0.58\\

UGC \phantom{0}7603 &\phantom{0}9.45& \phantom{0}0.80  &78&  1.24 & \phantom{0}8.24&1.04 & \phantom{0}0.41 &0.52 &
1.81\\

UGC \phantom{0}8490 &\phantom{0}5.28& \phantom{0}0.95  &50& 0.71 & 11.51 & 0.93 & \phantom{0}2.31 &2.44 &
1.88\\

UGC \phantom{0}9211 &14.70& \phantom{0}0.33  &44&  1.54 & \phantom{0}9.62 & 1.43 & \phantom{0}1.23 &3.69 &
1.55\\

UGC 11707 &21.46& \phantom{0}1.13&68&5.82 & 20.30 & 6.78 & \phantom{0}9.89 &8.76 &
1.77\\

UGC 11861 &19.55& \phantom{0}9.44 &50& 4.69 & 12.80 & 4.33 & 45.84 & 4.86 &
6.55\\

UGC 12060 &15.10& \phantom{0}0.39 &40& 1.70 &  \phantom{0} 9.89 & 1.67 & \phantom{0}3.45 &8.94 &
2.00\\

UGC 12632 &\phantom{0}9.20& \phantom{0}0.86  &46& 3.43 & 11.38 & 1.55 & \phantom{0}4.26 &4.97 &
1.82\\

UGC 12732 &12.40& \phantom{0}0.71  &39& 2.10 & 14.43 & 3.23 & \phantom{0}4.11 &5.76 &
2.40\\
\hline
\end{tabular}
\label{table:24dwarfs}
\end{table}

\begin{figure}[H]
\epsfig{file=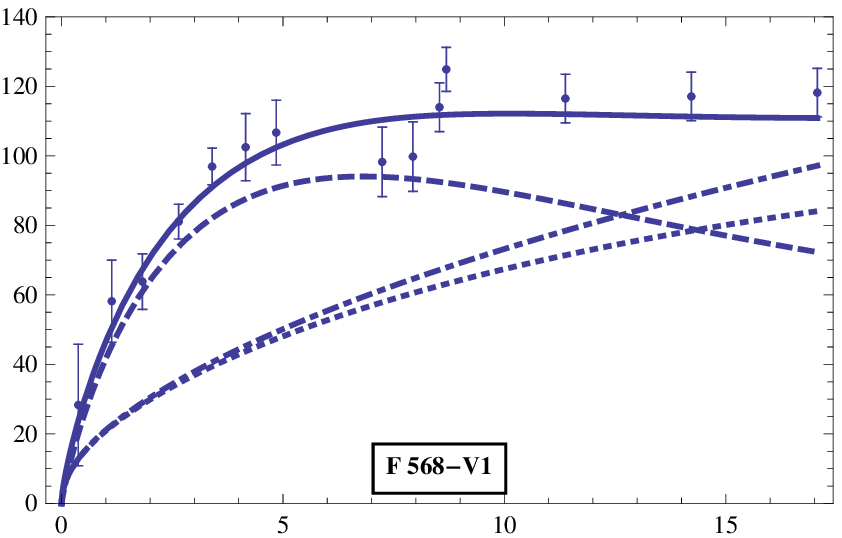,width=2.07in,height=1.8in}~~~
\epsfig{file=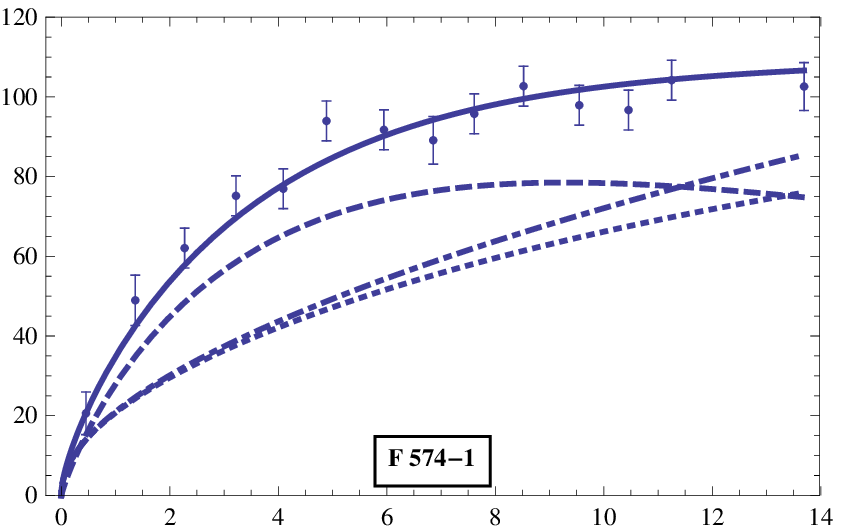,width=2.07in,height=1.8in}~~~
\epsfig{file=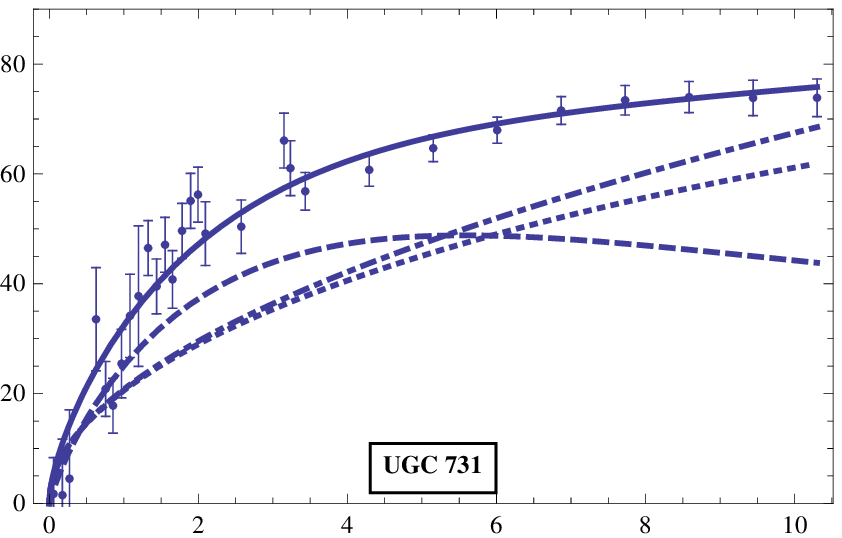,width=2.07in,height=1.8in}\\
\medskip
\epsfig{file=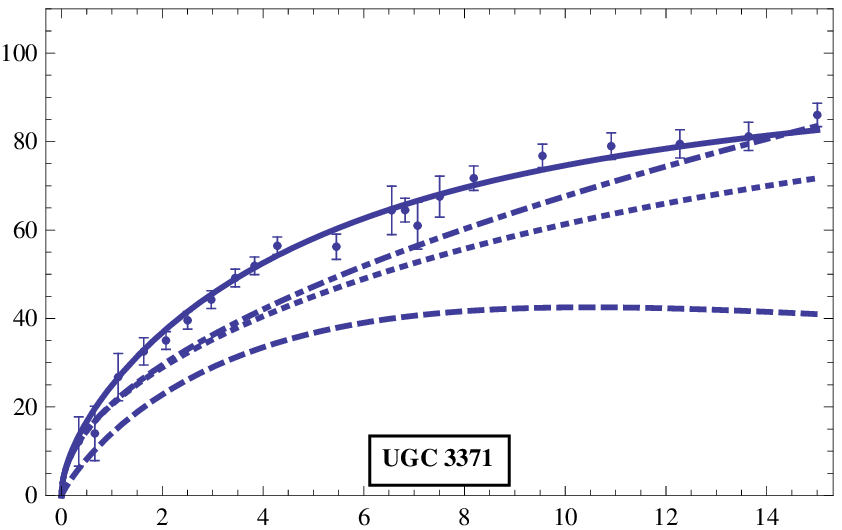,width=2.07in,height=1.8in}~~~
\epsfig{file=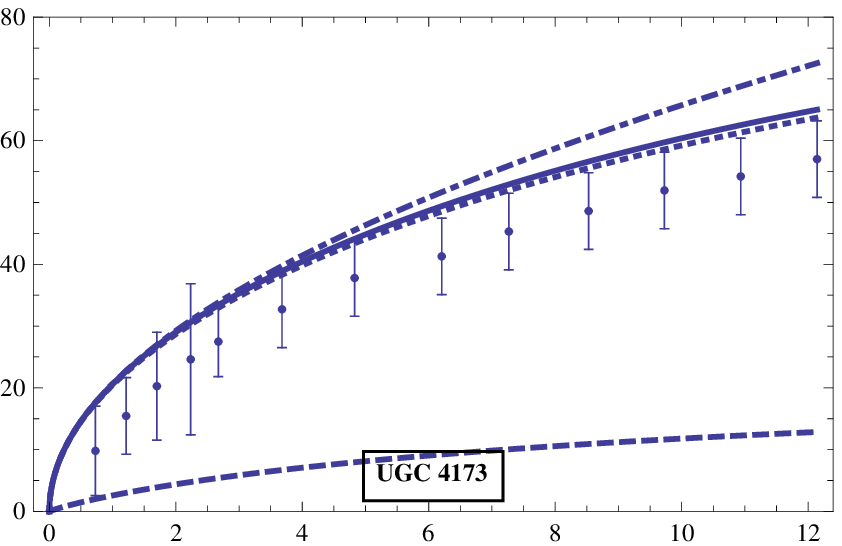,width=2.07in,height=1.8in}~~~
\epsfig{file=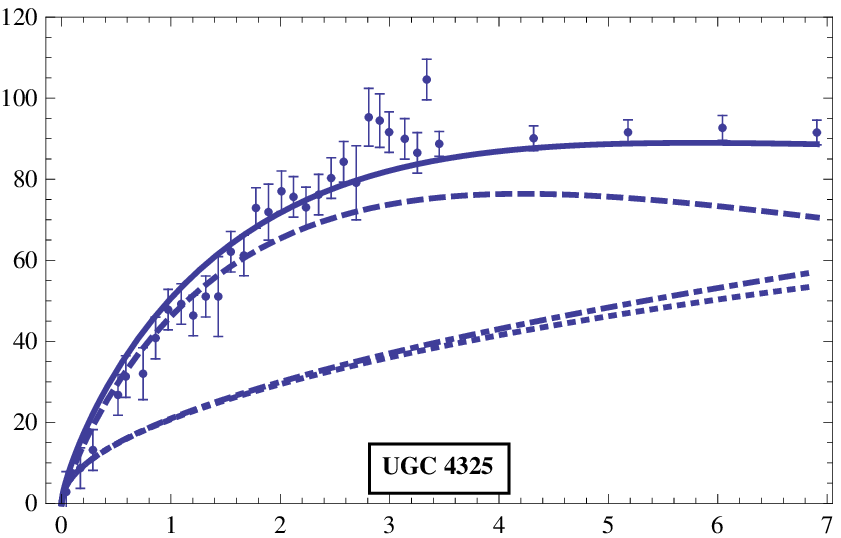,width=2.07in,height=1.8in}\\
\medskip
\epsfig{file=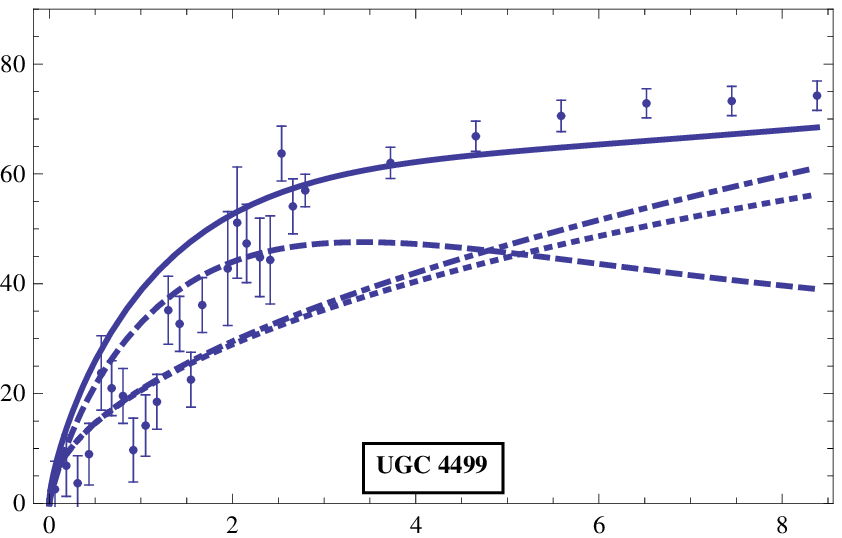,width=2.07in,height=1.8in}~~~
\epsfig{file=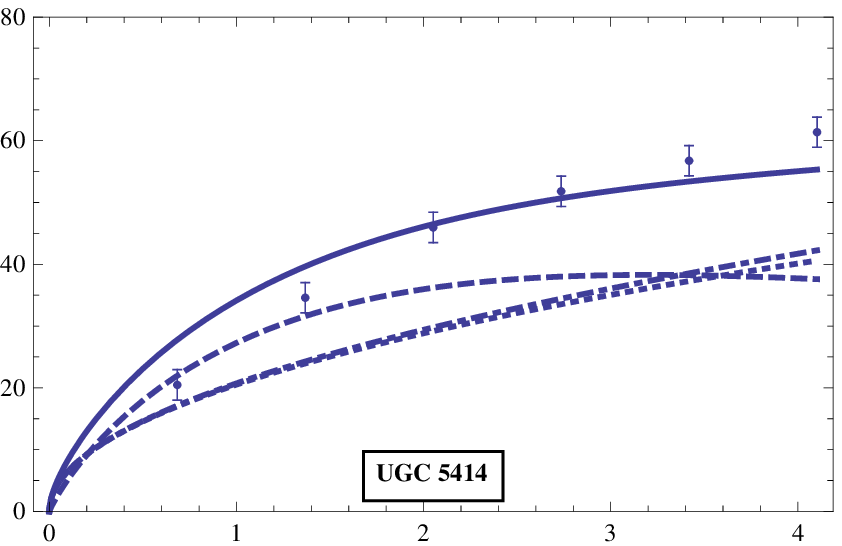,width=2.07in,height=1.8in}~~~
\epsfig{file=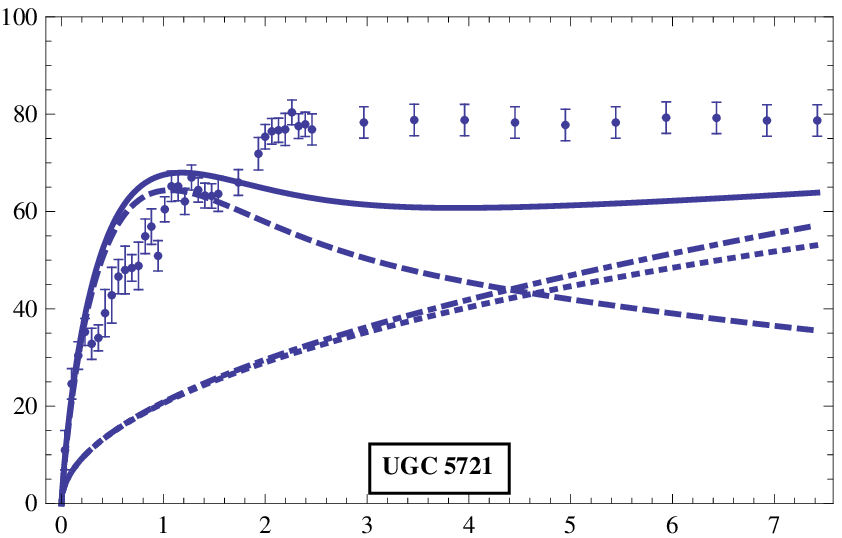,width=2.07in,height=1.8in}\\
\medskip
\epsfig{file=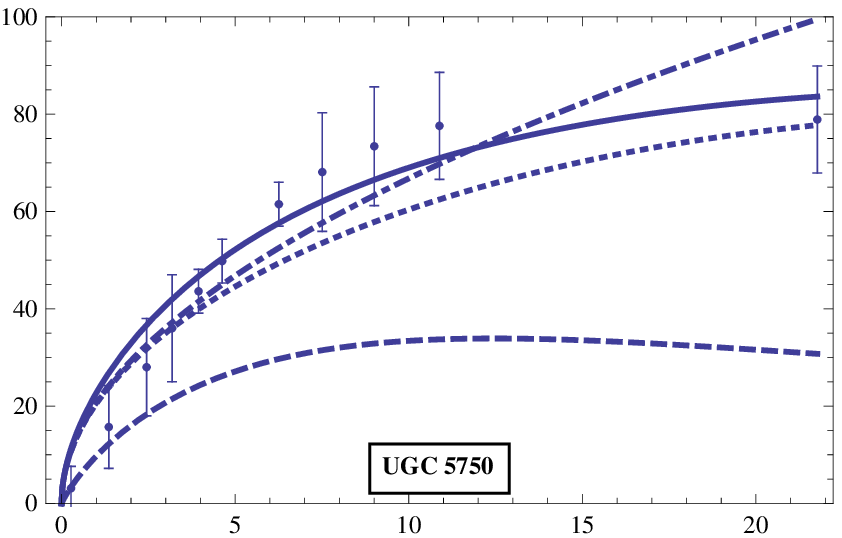,width=2.07in,height=1.8in}~~~
\epsfig{file=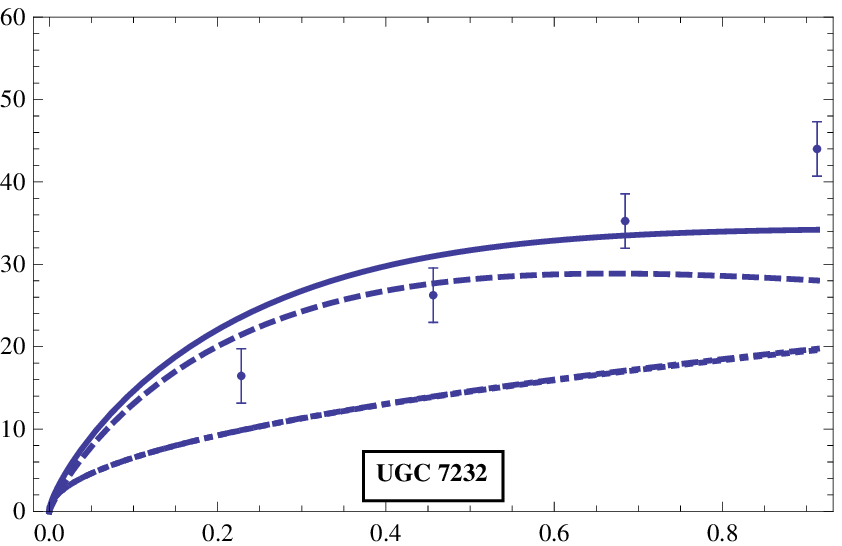,width=2.07in,height=1.8in}~~~
\epsfig{file=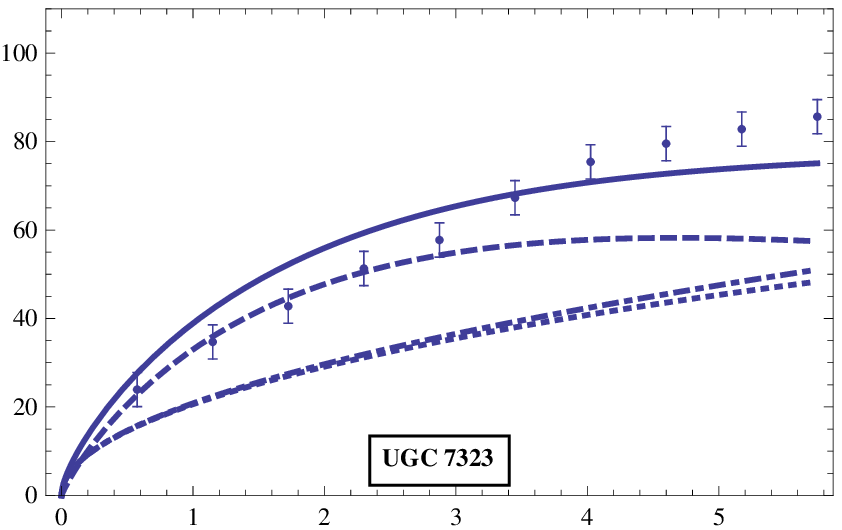,width=2.07in,height=1.8in}\\
\medskip
FIG.~1~PART 1:~Fitting to the rotational velocities (in ${\rm km}~{\rm sec}^{-1}$) of  the 24 galaxy sample with their quoted errors as plotted as a function of radial distance (in ${\rm kpc}$). For each galaxy we have exhibited the contribution due to the luminous Newtonian term alone (dashed curve), the contribution from the two linear terms alone (dot dashed curve), the contribution from the two linear terms and the quadratic term combined (dotted curve), with the full curve showing the total contribution. No dark matter is assumed.
\label{Fig. (1-1)}
\end{figure}

\begin{figure}[H]
\epsfig{file=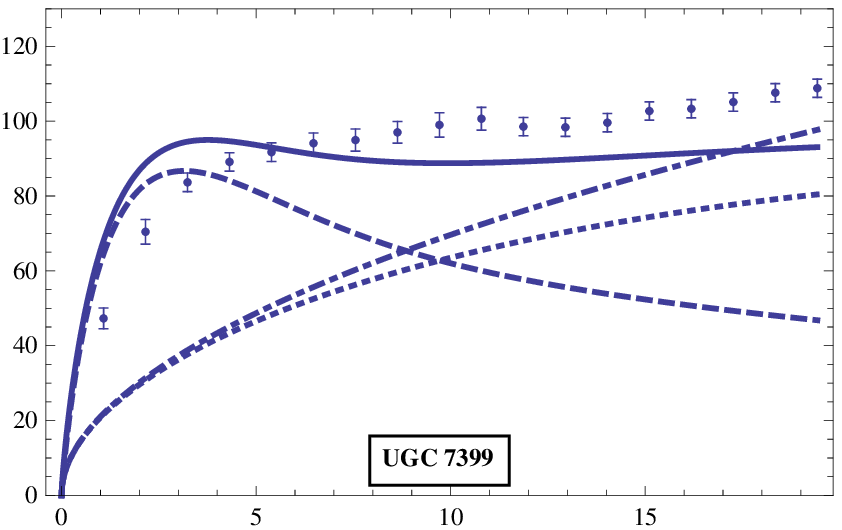,width=2.07in,height=1.8in}~~~
\epsfig{file=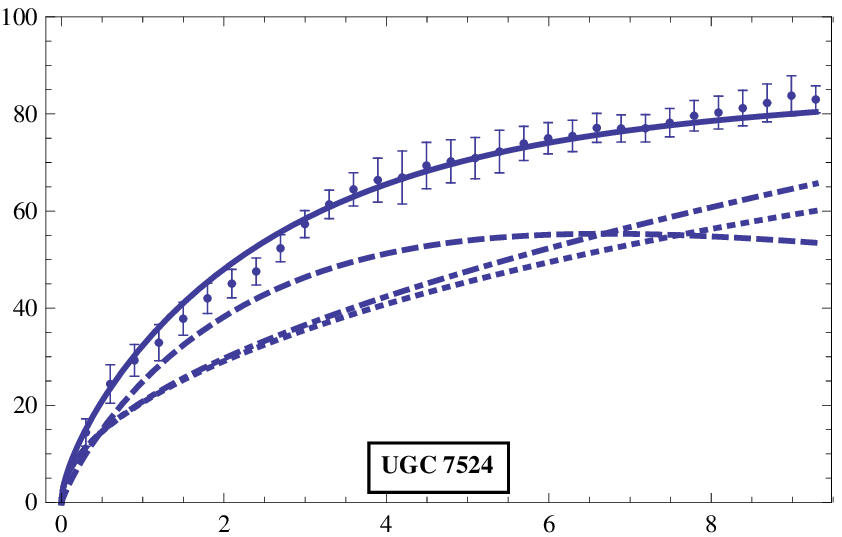,width=2.07in,height=1.8in}~~~
\epsfig{file=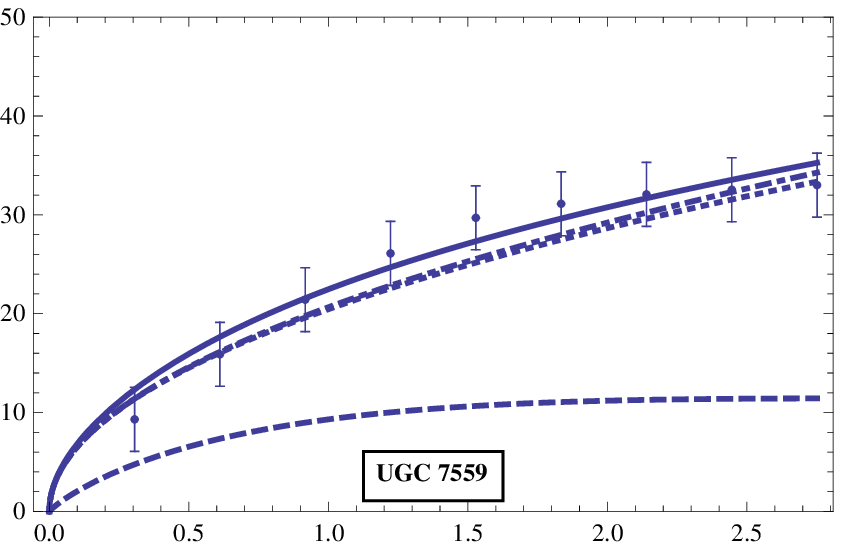,width=2.07in,height=1.8in}\\
\medskip
\epsfig{file=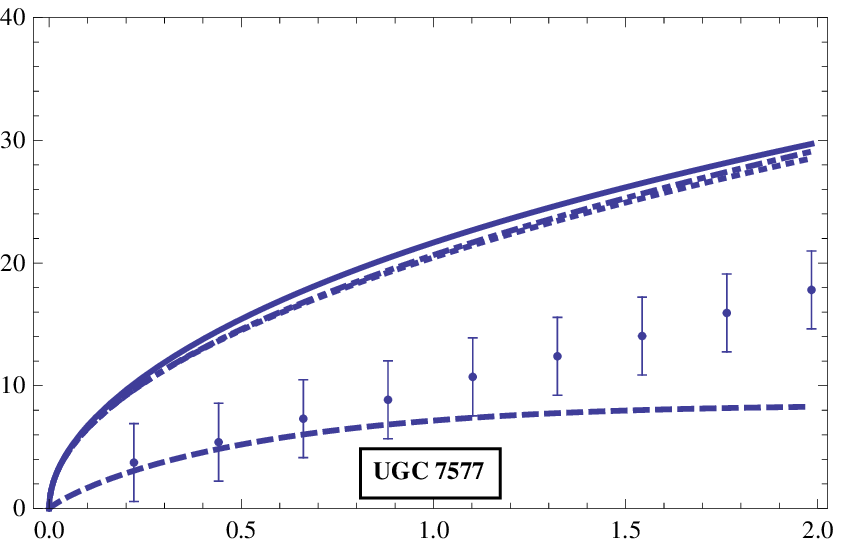,width=2.07in,height=1.8in}~~~
\epsfig{file=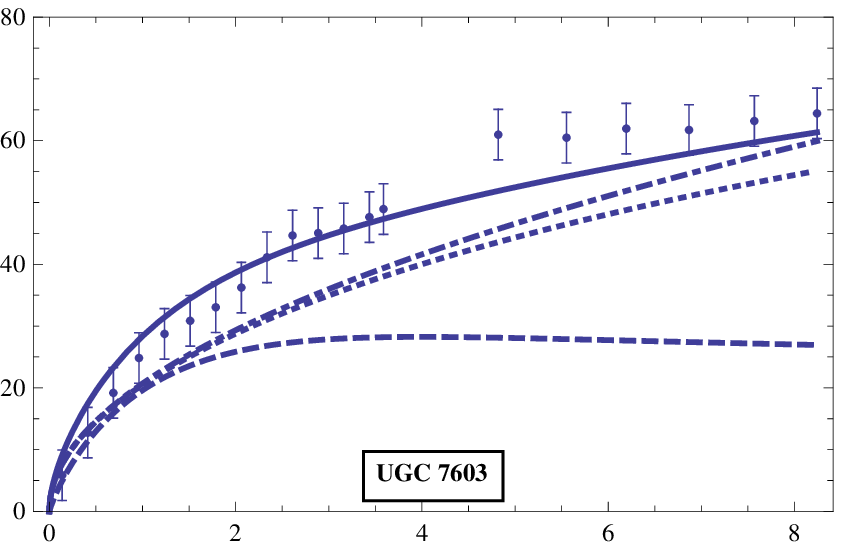,  width=2.07in,height=1.8in}~~~
\epsfig{file=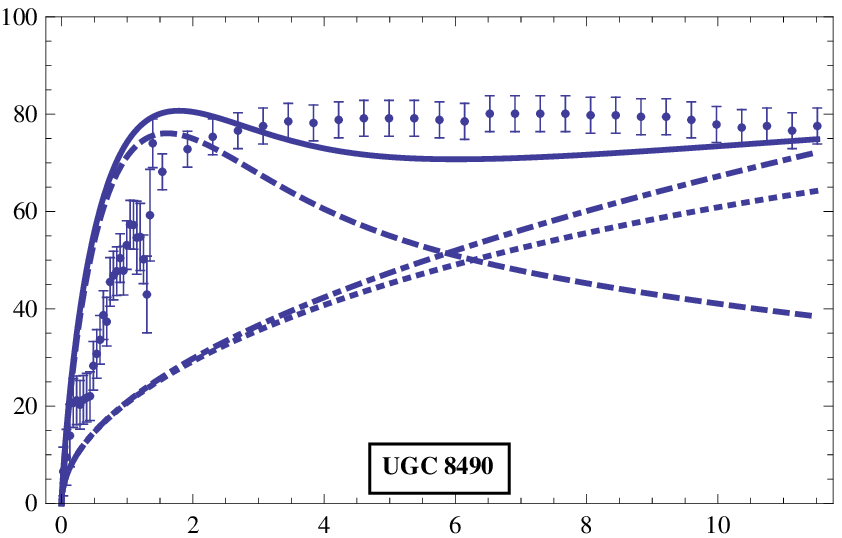,width=2.07in,height=1.8in}\\
\medskip
\epsfig{file=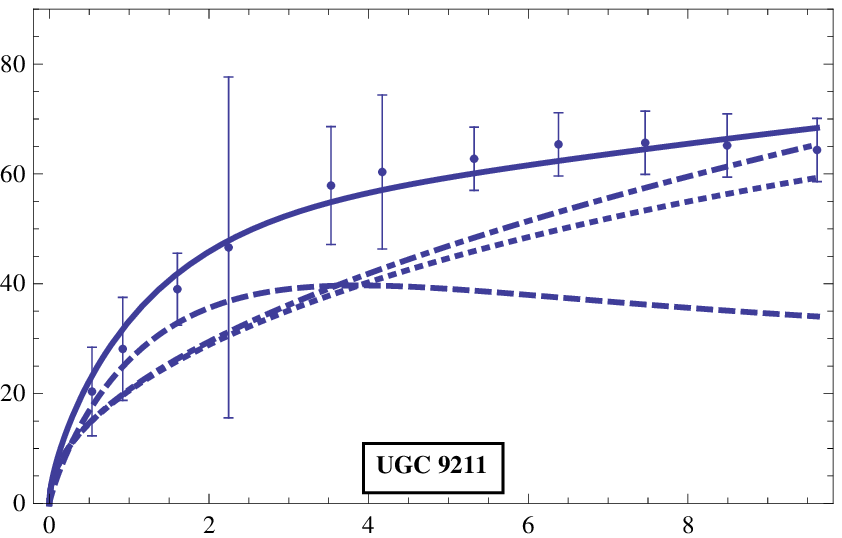,width=2.07in,height=1.8in}~~~
\epsfig{file=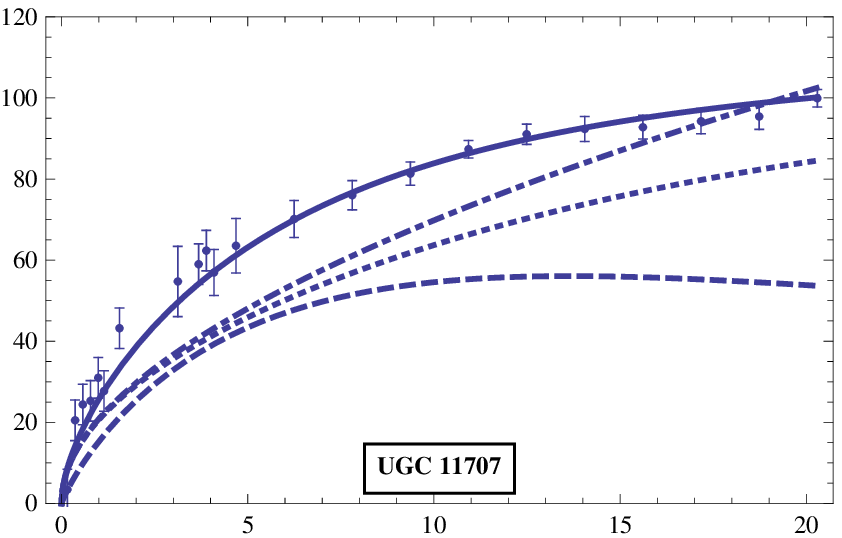,width=2.07in,height=1.8in}~~~
\epsfig{file=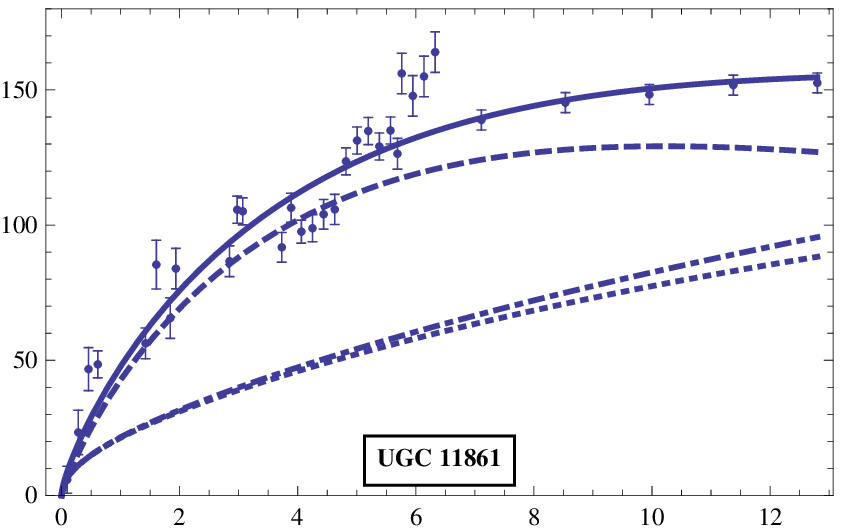,width=2.07in,height=1.8in}\\
\medskip
\epsfig{file=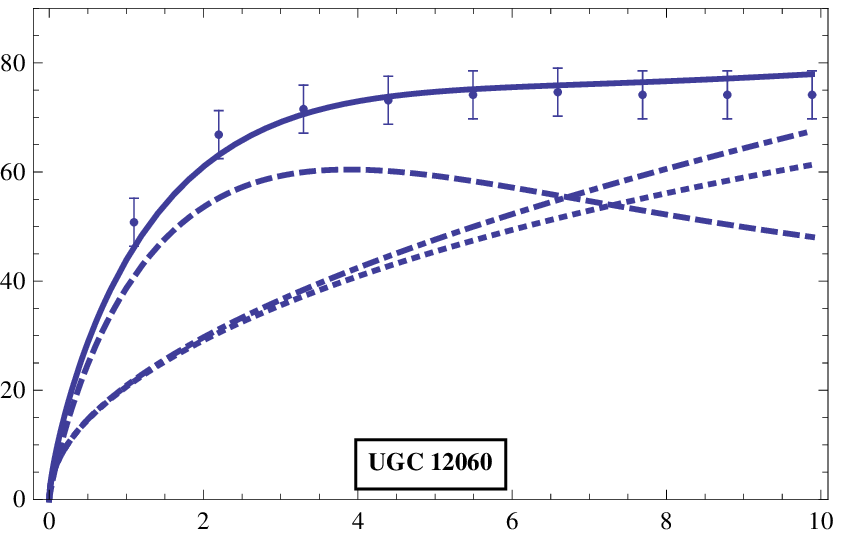,width=2.07in,height=1.8in}~~~
\epsfig{file=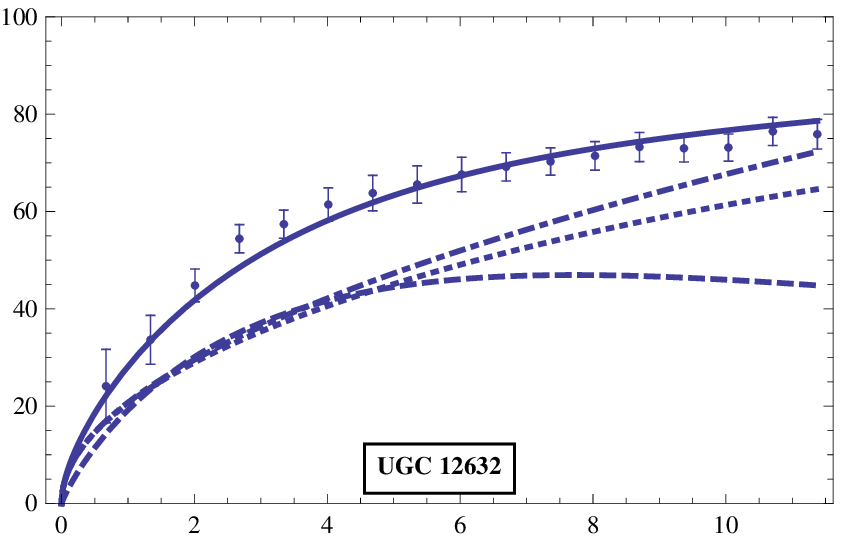,width=2.07in,height=1.8in}~~~
\epsfig{file=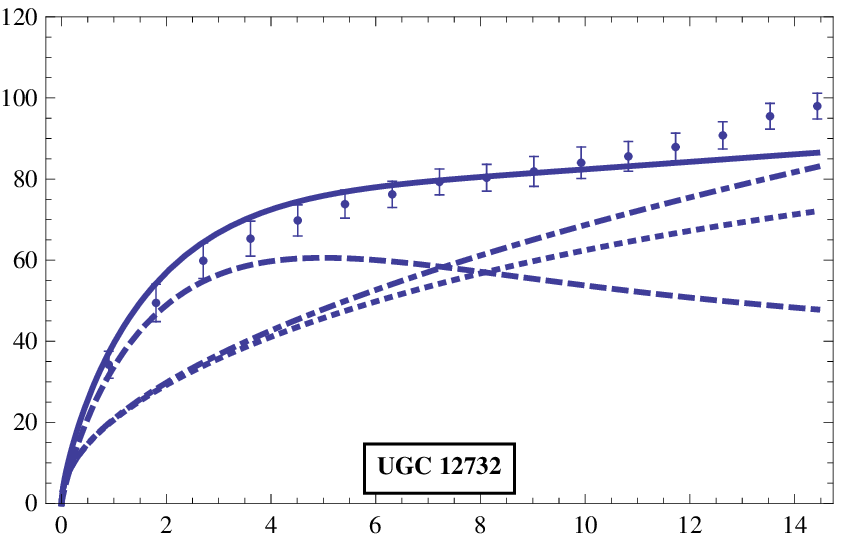,width=2.07in,height=1.8in}\\
\medskip
FIG.~1~PART 2:~Fitting to the rotational velocities of the 24 galaxy sample.
\label{Fig. (1-2)}
\end{figure}

\begin{table}[ht!]
\caption{Properties of the 6  Galaxy Sample}
\centering
\begin{tabular}{l c c c c c c c c c c} 
\hline\hline
\phantom{00}Galaxy\phantom{0}&\phantom{00}Distance  & $L_{\rm  B}$ & ~~$i$~~& $(R_0)_{\rm disk}$  & $R_{\rm last} $ &  $M_{\rm HI} $ & $M_{\rm disk}$ &  $ 
(M/L_{\rm B}) _{\rm disk}$ & $(v^2 / c^2 R)_{\rm last}$  \\  
&   (Mpc)  &  $(10^{9}{\rm L}_{\odot}^{\rm B})$  & $~~^{\circ}~~$ &  (kpc) & (kpc) & {$(10^{9} M_\odot)$} & {$(10^{9}
M_\odot)$} & ({$M_{\odot}/L_{\odot}^{\rm B}$}) & {$(10^{-30}\texttt{cm}^{-1})$} \\
\hline

UGC \phantom{0}3851 &\phantom{0}3.86& \phantom{0}1.48  & 63& 1.64 & \phantom{0}9.31 & 0.84 & \phantom{0}0.30 & 0.20 &
1.72  \\

UGC \phantom{0}4305 &\phantom{0}3.33& \phantom{0}0.83  & 50&  0.96& \phantom{0}6.75 & 0.57 & \phantom{0}0.17 & 0.20 &
0.71\\

UGC \phantom{0}4459 &\phantom{0}3.06& \phantom{0}0.03 &27  &0.60 & \phantom{0}2.47 & 0.04 & \phantom{0}0.01 &0.20&
0.97\\

UGC \phantom{0}5139 &\phantom{0}4.69& \phantom{0}0.20 &14&0.96 & \phantom{0}3.58& 0.21 & \phantom{0}0.05 &0.24 &
1.19\\

UGC \phantom{0}5423 &\phantom{0}7.14& \phantom{0}0.14  & 45 &0.61 & \phantom{0}1.97 & 0.05 & \phantom{0}0.28 &2.01 &
1.82\\

UGC \phantom{0}5666 &\phantom{0}3.85& \phantom{0}2.53  & 56 &3.56 & 11.25 & 1.37& \phantom{0}1.96 &0.77 &
1.88\\

\hline
\end{tabular}
\label{table:6dwarfs}
\end{table}

\begin{figure}
\epsfig{file=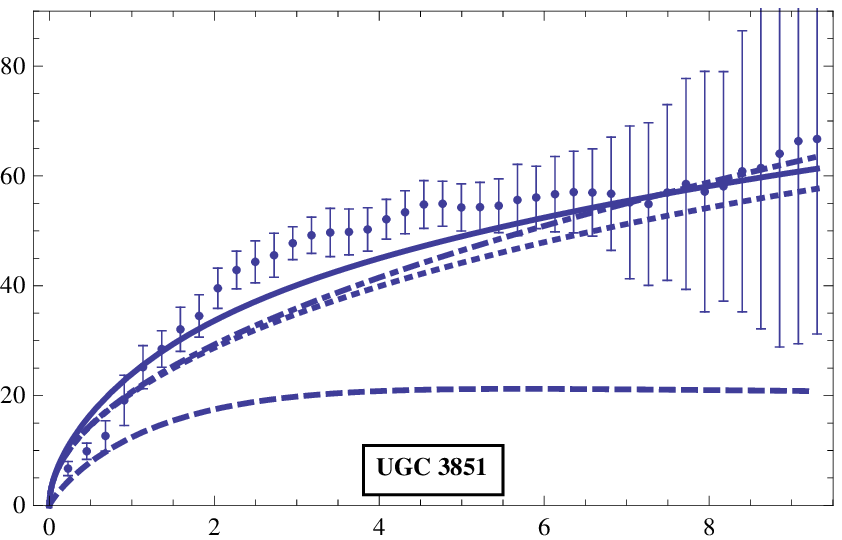,width=2.07in,height=1.8in}~~~
\epsfig{file=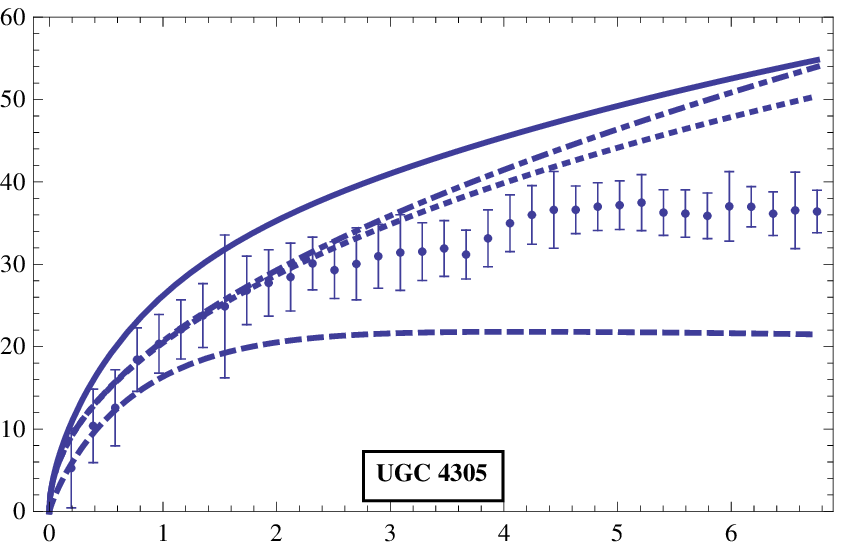,width=2.07in,height=1.8in}~~~
\epsfig{file=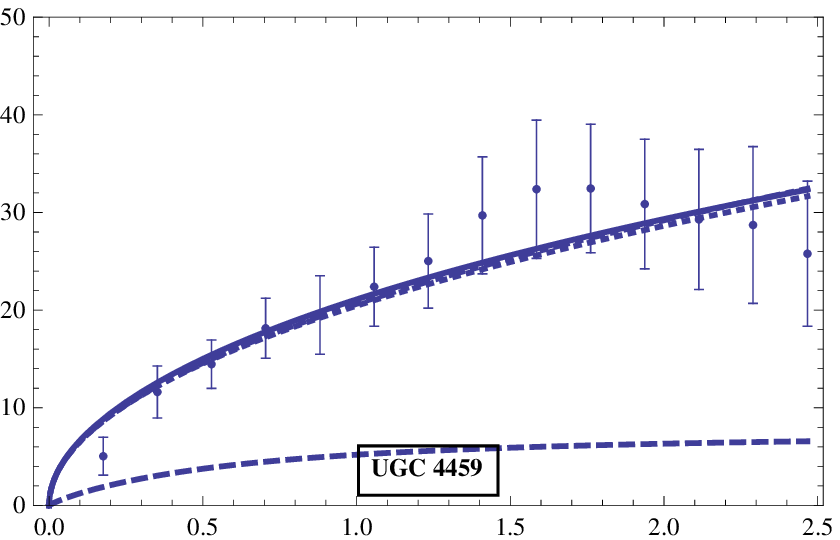,width=2.07in,height=1.8in}\\
\medskip
\epsfig{file=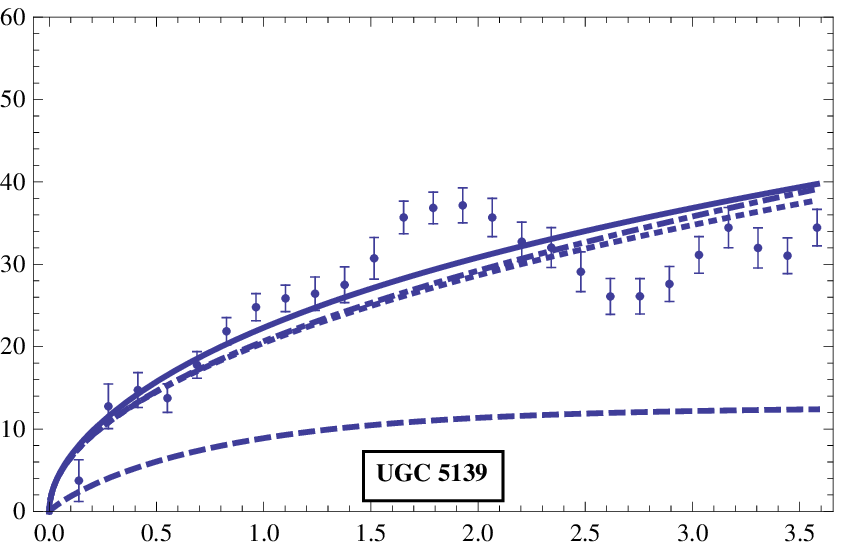,width=2.07in,height=1.8in}~~~
\epsfig{file=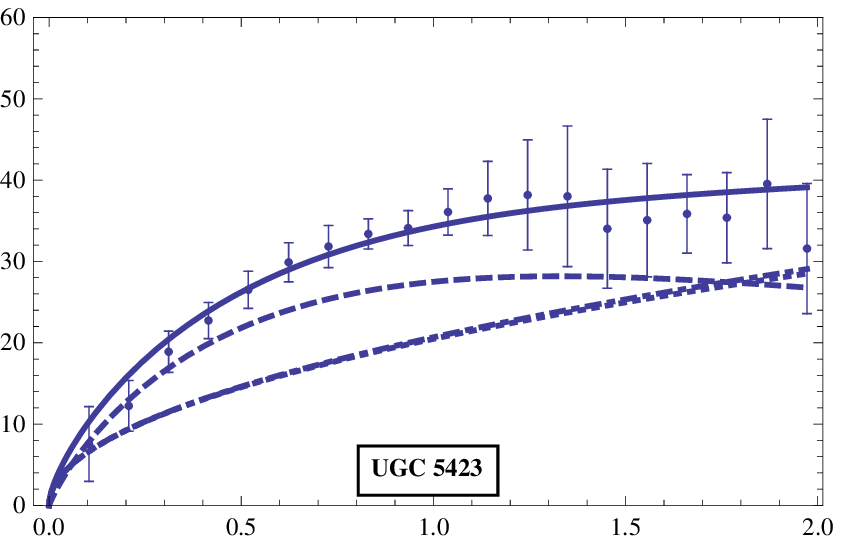,width=2.07in,height=1.8in}~~~
\epsfig{file=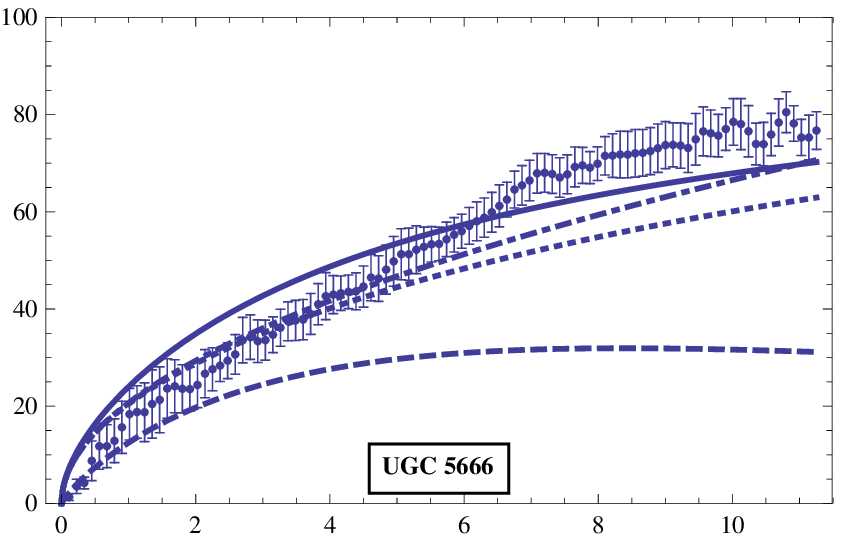,width=2.07in,height=1.8in}\\
\medskip
FIG.~2~Fitting to the rotational velocities of the 6 galaxy sample.
\label{Fig. (2)}
\end{figure}

\begin{table}[ht!]
\caption{Properties of the 7 Galaxies that Benefited from Parameter Adjustment}
\centering
\begin{tabular}{l c c c c c c c c c c} 
\hline\hline
\phantom{00}Galaxy\phantom{0}&\phantom{00}Distance  & $L_{\rm  B}$ & ~$\Delta i$~~& $(R_0)_{\rm disk}$  & $R_{\rm last} $ &  $M_{\rm HI} $ & $M_{\rm disk}$ &  $ 
(M/L_{\rm B}) _{\rm disk}$ & $(v^2 / c^2 R)_{\rm last}$  \\  
&   (Mpc)  &  $(10^{9}{\rm L}_{\odot}^{\rm B})$  & $~~^{\circ}~~$ & (kpc) & (kpc) & {$(10^{9} M_\odot)$} & {$(10^{9}
M_\odot)$} & ({$M_{\odot}/L_{\odot}^{\rm B}$}) & {$(10^{-30}\texttt{cm}^{-1})$} \\
\hline

UGC \phantom{0}3851 &\phantom{0}4.85& \phantom{0}2.33  & & 2.07 &11.70 & 1.32 & \phantom{0}0.47 & 0.20 &
1.37 \\

UGC \phantom{0}4305 &\phantom{0}2.34& \phantom{0}0.41  &~\phantom{0}--5&  0.68& \phantom{0}4.75 & 0.31 & \phantom{0}0.08 & 0.20 &
1.18\\

UGC \phantom{0}4173 &16.70& \phantom{0}0.33  &~\phantom{0}--5& 4.43 & 12.14 & 2.24 & \phantom{0}0.07 & 0.20 &
1.21  \\

UGC \phantom{0}5721 &\phantom{0}7.60& \phantom{0}0.48  & +10&  0.76 & \phantom{0}8.41 & 0.57 & \phantom{0}1.90 & 3.96 &
2.27\\

UGC \phantom{0}7399 &24.66& \phantom{0}4.61 &  &2.32 & 32.30 & 6.38 & \phantom{0}5.42 &1.18 &
1.32\\

UGC \phantom{0}7577 &\phantom{0}2.13& \phantom{0}0.05 &~--15&0.51 & \phantom{0}1.39 & 0.04 & \phantom{0}0.01 &0.20 &
1.18\\

UGC \phantom{0}8490 &\phantom{0}5.28& \phantom{0}0.95  & +10 &0.71 & 11.51 & 0.72 & \phantom{0}1.53 &1.61 &
1.47\\

\hline
\end{tabular}
\label{table:dwarfsadjusted}
\end{table}

\begin{figure}
\epsfig{file=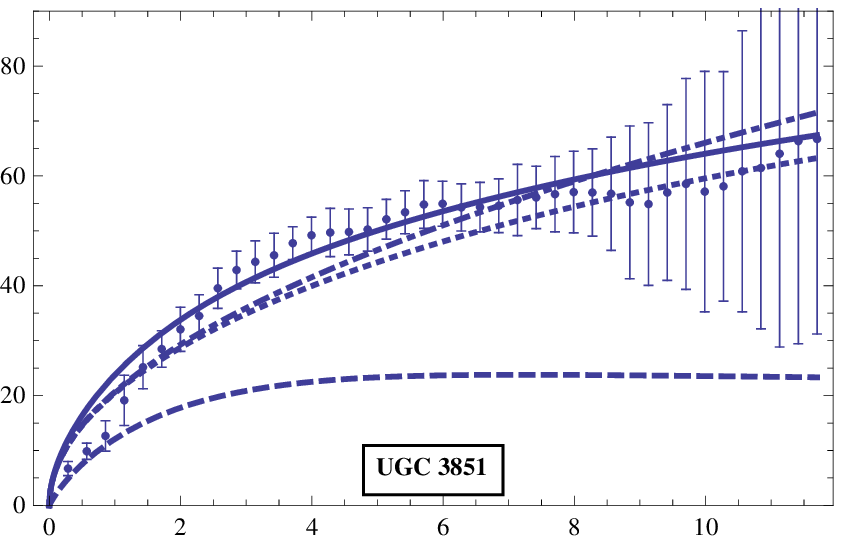,width=2.07in,height=1.65in}~~~
\epsfig{file=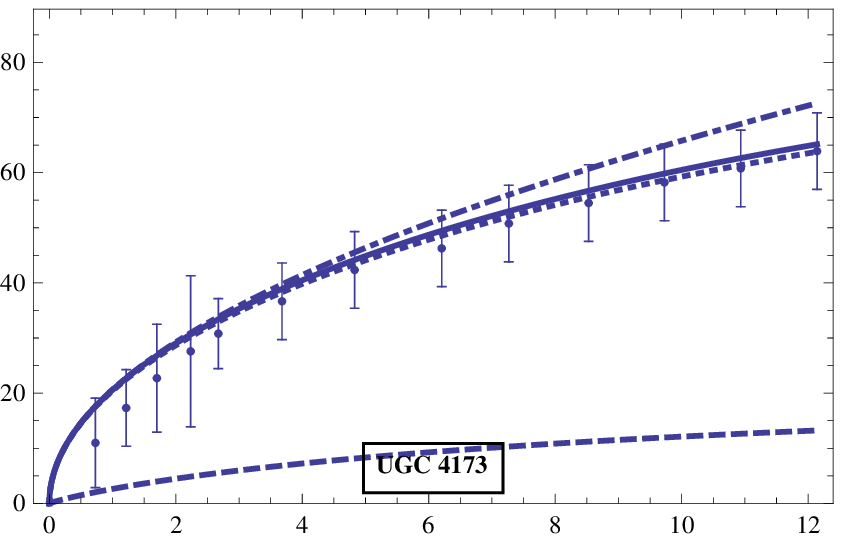,width=2.07in,height=1.65in}~~~
\epsfig{file=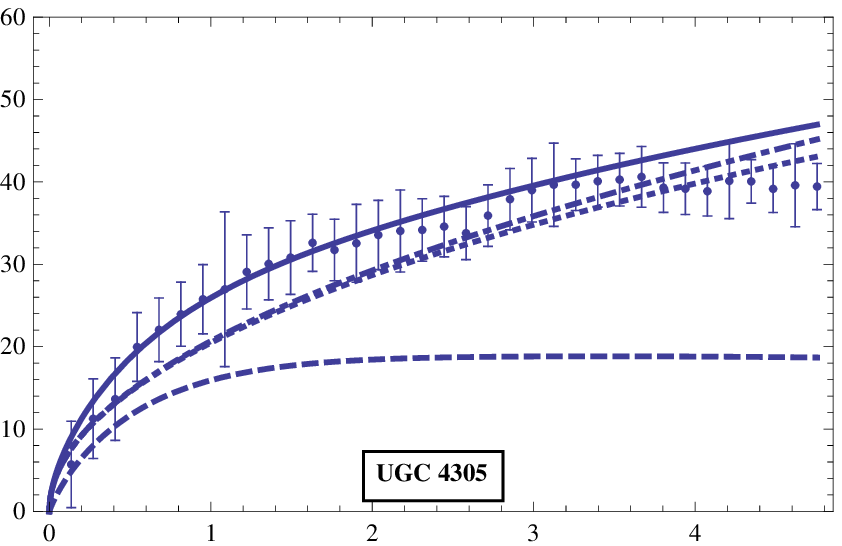,width=2.07in,height=1.65in}\\
\medskip
\epsfig{file=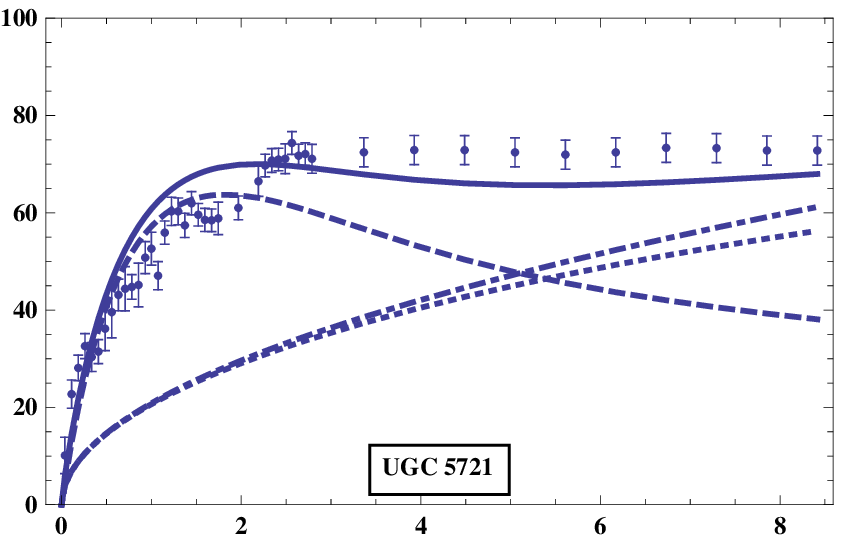,width=2.07in,height=1.65in}~~~
\epsfig{file=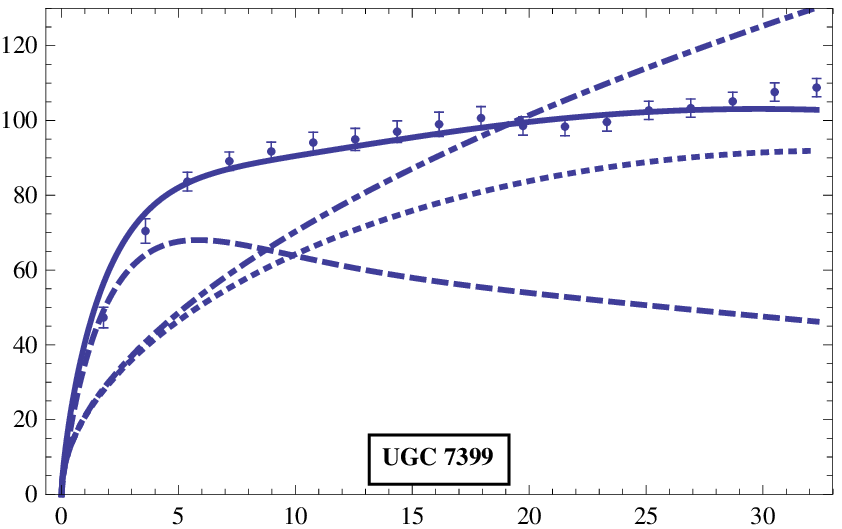,width=2.07in,height=1.65in}~~~
\epsfig{file=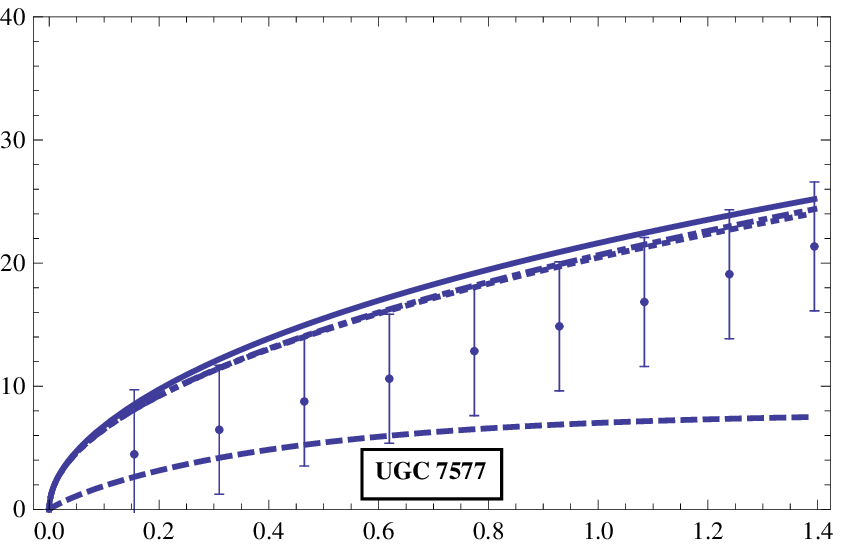,width=2.07in,height=1.65in}\\
\medskip
\epsfig{file=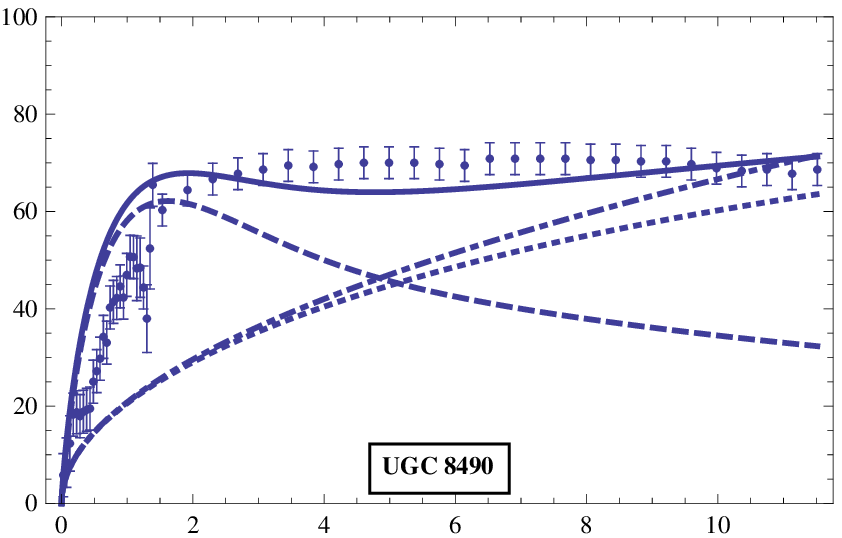,width=2.07in,height=1.65in}\\
\medskip
FIG.~3~Fitting to the rotational velocities of the 7 galaxies  that benefited from parameter adjustment.
\label{Fig. (3)}
\end{figure}


\begin{thebibliography}{}



\bibitem{Mannheim2006} P.~D.~Mannheim,~Prog.~Part.~Nucl.~Phys.~{\bf 56},~340~(2006). 

\bibitem{Mannheim2009} P.~D.~Mannheim,~Gen.~Rev.~Gravit.~{\bf 43}, 703 (2011).

\bibitem{Bender2008} C.~M.~Bender and P.~D.~Mannheim, Phys.~Rev.~Lett.~{\bf 100},~110402 (2008);~Phys.~Rev.~D {\bf 78}, 025022 (2008).

\bibitem{Mannheim2011} P.~D.~Mannheim,~{\it Making the case for conformal gravity}, January 2011,  Found.~Phys.~in press (arXiv:1101.2186 [hep-th]). Presentation at the International Conference on Two Cosmological Models, Universidad Iberoamericana, Mexico City, November 2010.

\bibitem{Mannheim1997} P.~D.~Mannheim,~Ap.~J.~{\bf 479}, 659 (1997).

\bibitem{Mannheim2010a} P.~D.~Mannheim and J.~G.~O'Brien,~Phys.~Rev.~Lett.~{\bf 106}, 121101 (2011).

\bibitem{Mannheim2010b} P.~D.~Mannheim and J.~G.~O'Brien,~{\it Fitting galactic rotation curves with conformal gravity and a global quadratic potential},~November 2010. (arXiv:1011.3495 [astro-ph.CO])

\bibitem{Swaters2009} R.~A.~Swaters,~R.~Sancisi,~T.~S.~van Albada~and~J.~M.~van der Hulst, A.~A.~{\bf 493}, 871 (2009).

\bibitem{Oh2011} S-H.~Oh, W.~J.~G.~de Blok, E.~Brinks, F.~Walter, R.~C.~Kennicutt,~A.~J.~{\bf 141}, 193 (2011).

\bibitem{Mannheim1989} P.~D.~Mannheim and D.~Kazanas,~Ap.~J.~{\bf 342}, 635 (1989).

\bibitem{Mannheim1994} P.~D.~Mannheim and D.~Kazanas,~Gen.~Rel.~Gravit.~{\bf 26}, 337 (1994).

\bibitem{Swaters2010} R.~A.~Swaters, R.~H.~Sanders and S.~S.~McGaugh, Ap.~J.~{\bf 718}, 380 (2010). 

\bibitem{Swaters1999} R.~A.~Swaters, Ph.~D.~Dissertation, Rijksuniversiteit Groningen (1999).

\bibitem{Swaters2003} R.~A.~Swaters, B.~F.~Madore, F.~C.~van den Bosch and M.~ Balcells, Ap.~J.~{\bf 583}, 732 (2003).



\bibitem{deBlok2001}  W.~J.~G.~de Blok, S.~S.~McGaugh and V.~C.~Rubin, A.~J.~{\bf 122}, 2396 (2001).


\bibitem{Swaters2002a} R.~A.~Swaters and M.~Balcells, A.~A.~{\bf 390}, 863 (2002). 

\bibitem{Swaters2002b} R.~A.~Swaters, T.~S.~van Albada, J.~M.~van der Hulst and R.~ Sancisi, A.~A.~{\bf 390}, 829 (2002). 

\bibitem{deBlok1997} W.~J.~G.~de Blok and S.~S.~McGaugh, Mon.~Not.~R.~Astron.~Soc.~{\bf 290}, 533 (1997).


\bibitem{Walter2008} F.~Walter, E.~Brinks, W.~J.~G.~de Blok, F.~Bigiel, R.~C.~Kennicutt, M.~D.~Thornley and A.~Leroy, A.~J.~{\bf 136},  2563 (2008).



\bibitem{Pasquali2008} A.~Pasquali, A.~Leroy, H.-W.~Rix, F.~Walter, T.~Herbst, E.~Giallongo, R.~Ragazzoni, A.~Baruffolo, R.~Speziali, J.~Hill, G.~Beccari, N.~Bouche, P.~Buschkamp, C.~Kochanek, E.~Skillman and J.~Bechtold, Ap.~J.~{\bf  687}, 1004 (2008).

\bibitem{Spano2008} M.~Spano, M.~Marcelin, P.~Amram, C.~Carignan, B.~Epinat, O.~Hernandez,
Mon.~Not.~R.~Astron.~Soc.~{\bf 383}, 297 (2008).


\bibitem{Mannheim1990} P.~D.~Mannheim,~Gen.~Rel.~Gravit.~{\bf 22}, 289 (1990).




\bibitem{Milgrom1983} M.~Milgrom,~Ap.~J.~ {\bf 270}, 365, 371, 384 (1983).







\bibitem{Moffat2005} J.~W.~Moffat, J.~Cos.~Ast.~Phys.~{\bf 05},~003 (2005);~{\bf 03},~004 (2006); J.~R.~Brownstein and  J.~W.~Moffat,~Ap.~J.~{\bf 636}, 721 (2006).


\bibitem{Navarro1996} J. F. Navarro, C. S. Frenk and S. D. M. White,~Ap.~J.~{\bf 462}, 563 (1996); {\bf 490}, 493 (1997).


\end{thebibliography}
\end{document}